\documentclass{aa}

\usepackage{graphicx}
\usepackage{txfonts}
\usepackage{url}
\usepackage{natbib}
\usepackage{color}
\bibpunct{(}{)}{;}{a}{}{,}
\begin{document}
\title{\textcolor{white}{b} Mantle formation, coagulation and the origin of cloud/core-shine \textcolor{white}{b} II. Comparison with observations}

\author{N. Ysard\inst{\ref{inst1}}
\and M. K\"ohler\inst{\ref{inst1}, \ref{inst2}}
\and A. Jones\inst{\ref{inst1}}
\and E. Dartois\inst{\ref{inst1}}
\and M. Godard\inst{\ref{inst3}}
\and L. Gavilan\inst{\ref{inst1}}}

\institute{Institut d'Astrophysique Spatiale, UMR8617, CNRS/Universit{\'e} Paris Sud, Universit{\'e} Paris Saclay, Universit{\'e} Paris Sud, F-91400 Orsay, France\label{inst1}
\and School of Physics and Astronomy, Queen Mary University of London, 327 Mile End Road, London, E1 4NS, UK\label{inst2}
\and Centre de Sciences Nucl{\'e}aires et de la Mati{\`e}re (CSNSM), UMR9609, CNRS/Universit{\'e} Paris Sud, Universit{\'e} Paris Saclay, Universit{\'e} Paris Sud, F-91400 Orsay, France\label{inst3} \textcolor{white}{aaaaaaaaaaaaaaaaaaaaaaaaaaaaaaaaaaaaaaaaaaaaaaaaaaaaaaaaaaaaaaaaaaaaaa aaaaa} \email{nathalie.ysard@ias.u-psud.fr}}

\abstract{Many dense interstellar clouds are observable in emission in the near-IR (J, H, and K photometric bands), commonly referred to as ``Cloud-shine'', and in the mid-IR (Spitzer IRAC 3.6 and 4.5~$\mu$m bands), the so-called ``Core-shine''. These C-shine observations have usually been explained in terms of grain growth but no model has yet been able to self-consistently explain the dust spectral energy distribution from the near-IR to the submm.}
{We want to demonstrate the ability of our new core/mantle evolutionary dust model THEMIS (The Heterogeneous dust Evolution Model at the IaS), which has been shown to be valid in the far-IR and submm, to reproduce the C-shine observations.}
{Our starting point is a physically motivated core/mantle dust model. It consists of three dust populations: small poly-aromatic-rich carbon grains ; bigger core/mantle grains with mantles of aromatic-rich carbon and cores either made of amorphous aliphatic-rich carbon or amorphous silicate. Then, we assume an evolutionary path where these grains, when entering denser regions, may first form a second aliphatic-rich carbon mantle (coagulation of small grains, accretion of carbon from the gas phase), second coagulate together to form large aggregates, and third accrete gas phase molecules coating them with an ice mantle. To compute the corresponding dust emission and scattering, we use a 3D Monte-Carlo radiative transfer code.}
{We show that our global evolutionary dust modelling approach THEMIS allows us to reproduce C-shine observations towards dense starless clouds. Dust scattering and emission is most sensitive to the cloud central density and to the steepness of the cloud density profile. Varying these two parameters leads to changes, which are stronger in the near-IR, in both the C-shine intensity and profile.}
{With a combination of aliphatic-rich mantle formation and low-level coagulation into aggregates, we can self-consistently explain the observed C-shine and far-IR/submm emission towards dense starless clouds.}

   \authorrunning{}
\titlerunning{Mantle formation, coagulation and the origin of cloud/core shine: II. Comparison with observations}

\maketitle
%

\section{Introduction}
\label{introduction}

Variations in the dust spectral energy distribution (SED) from the diffuse interstellar medium (ISM) to dense molecular clouds have clearly been observed in the Milky Way. These variations encompass variations in the dust temperature, opacity, and spectral index as measured from dust thermal emission in the far-IR and submm \citep[][among many others]{Lagache1998, Stepnik2003, Ridderstad2006, Schnee2010, Juvela2012, Roy2013}, as well as variations in the mid- to far-IR intensity ratio \citep[see for instance][]{Laureijs1991, Bernard1999, Stepnik2003}. This is usually interpreted in terms of grain evolution through grain growth and ice mantle formation with increasing local density \citep{Boogert2015, Koehler2015, Ormel2009}. Spectroscopically, an excess absorption in the red wing of the water ice mantle band observed at $3~\mu$m in dense clouds may be associated with grain growth to bigger sizes than those expected in the diffuse ISM \citep[e.g.][]{Smith1989}. Such bigger grains were also suggested as potentially contributing to the long-wavelength wing of the $4.67~\mu$m CO ice band \citep{Dartois2006}. Apart from its IR to submm thermal emission, dust evolution was also long ago shown by scattered light at shorter wavelengths. First observed in the visible \citep{Struve1936, Struve1937, Witt1968, Mattila1970a, Mattila1970b}, dust scattering was then measured in the near-IR \citep{Witt1994, Lehtinen1996, Nakajima2003, Foster2006, Juvela2008, Juvela2009, Andersen2013, Malinen2013} and in the mid-IR \citep{Pagani2010, Paladini2014}. Most of these studies concluded that dust grains with the same size distribution as in the diffuse ISM could not account for the measured brightness and thus that grain growth was required in dense clouds \citep[see for instance][]{Witt1994, Nakajima2003, Foster2006, Andersen2013, Lefevre2014,  Steinacker2010, Steinacker2014a, Steinacker2014b, Steinacker2015}. Dust emission observations in the near- and mid-IR were recently labelled cloud-shine \citep[][observations in the J, H, and K photometric bands]{Foster2006} and core-shine \citep[][in the L and M photometric bands]{Pagani2010}, respectively, and in the following we refer globally to these observations as C-shine since they both result from light scattering by dust grains. Even if there is a consensus about the need for grain growth in dense molecular clouds to explain C-shine, no model has yet been able to self-consistently explain the dust observations from the near-IR to the far-IR/submm.

A core-mantle dust model taking into account dust evolution both in the diffuse and dense ISM was recently developed \citep{Jones2012a, Jones2012b, Jones2012c, Jones2013, Koehler2014, Ysard2015, Koehler2015} and in the following this global evolutionary dust modelling approach is referred to as THEMIS (The Heterogeneous dust Evolution Model at the IaS). The ability of the THEMIS approach to reproduce the variations in the dust spectral energy distribution in terms of temperature, far-IR/submm opacity and spectral index, observed in both diffuse \citep{Ysard2015, Fanciullo2015} and dense ISM \citep{Koehler2015}, was previously demonstrated together with an overall agreement with extinction measurements. \citet{Jones2015}, hereafter paper I, investigated the scattering properties of THEMIS dust particles, showing that the formation of an aliphatic-rich carbon mantle onto diffuse-ISM type grains \citep{Jones2013, Koehler2014} decreases the dust absorption cross-section, while leaving the scattering cross-section unchanged, in the near- and mid-IR (see their Figs.~10 and 11). These changes lead to higher albedo dust in agreement with the observational results for dark clouds presented in \citet{Mattila1970b, Mattila1970a}, \citet{Witt1994}, and \citet{Lehtinen1996}. The main conclusion of paper I is that the evolution of dust through aliphatic-rich mantle formation\footnote{More generally, the formation of any material having a refractive index with a low imaginary part, $k$, would result in increasing the grain albedo.} and grain coagulation appears to be a likely explanation for the observed C-shine. The aim of the present study is to investigate to what extent THEMIS could account for the near- and mid-IR observations towards dense clouds.

This paper is organised as follows. Section~\ref{models} describes THEMIS and the tools used to calculate dust emission and scattering. Section~\ref{coreshine_cloudshine} explores the effect of changing various parameters describing the cloud structure and density, as well as parameters describing the radiation field on the dust scattering spectrum. In Sect.~\ref{comparison_with_observations}, the model results are compared to observations of dense molecular clouds in the near- and mid-IR. In Sect.~\ref{conclusions}, the astrophysical implications of our model results are discussed and concluding remarks given.

\section{Models and tools}
\label{models}

The interpretation of dust emission and scattering from ISM clouds depends on the dust model, the cloud structure, and the incident radiation field. Here, we first describe our view of dust evolution from the diffuse ISM to dense molecular clouds and the corresponding model characteristics (THEMIS, Sect.~\ref{dust_model}). Second, we detail the assumptions made to mimic the cloud geometry and the radiation field, and finally how the radiative transfer calculations are performed (Sect.~\ref{modelling_tools}). 

\begin{figure}[!t]
\centerline{\includegraphics[width=0.42\textwidth]{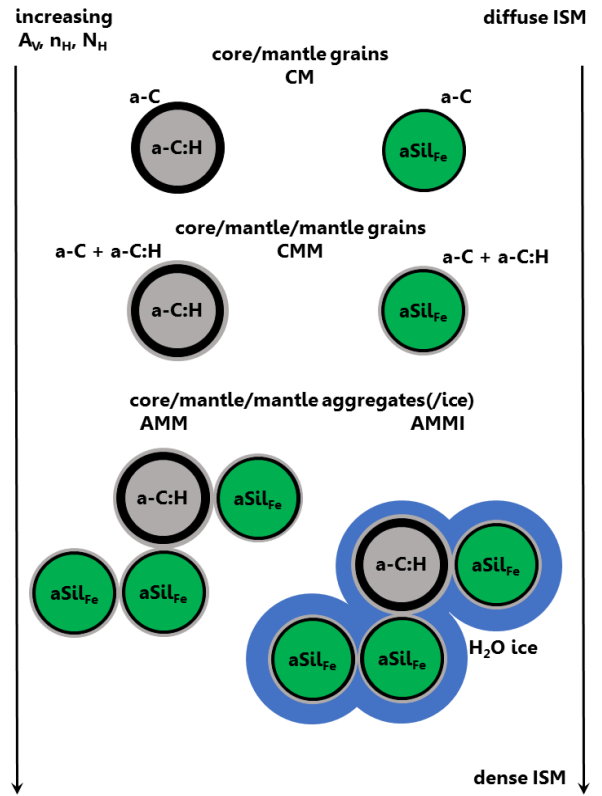}}
\caption{A schematic view of the dust composition and stratification from the diffuse ISM to dense molecular clouds. The gas density increases from the top to the bottom of the figure.}
\label{Fig1} 
\end{figure}

\subsection{THEMIS: from diffuse ISM to molecular clouds}
\label{dust_model}

Our starting point for the diffuse ISM is the \citet{Jones2013} dust model as updated by \citet{Koehler2014}. According to this model, grains up to 20~nm consist purely of aromatic-rich H-poor amorphous carbon, a-C, whereas bigger grains have a core/mantle (CM) structure, where the core consists either of amorphous silicate (forsterite and enstatite-normative compositions, Mg-rich) or of aliphatic-rich H-rich amorphous carbon, a-C:H. For both core types, the mantle consists of H-poor amorphous carbon. In this model, 199 ppm of C, 38.5 ppm of Si, 54.5 ppm of Mg, 131.5 ppm of O, 25.2 ppm of Fe, and 5.4 ppm of S are in grains \citep{Ysard2015}. Details of the model and optical property calculations can be found in \citet{Jones2012a, Jones2012b, Jones2012c}, \citet{Jones2013}, and \citet{Koehler2014}.

\begin{figure}[!t]
\centerline{\includegraphics[width=0.4\textwidth]{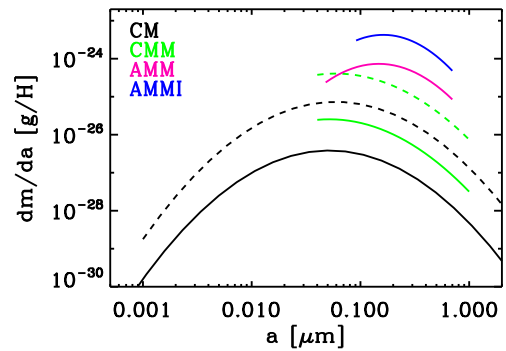}}
\caption{Mass distribution of THEMIS dust populations: the biggest grains of the CM populations are in black (solid line for the carbonaceous grains and dashed line for silicates), CMM populations in green (same linestyle as for CM grains), AMM in magenta, and AMMI in blue.}
\label{Fig2} 
\end{figure}

As described in \citet{Koehler2015}\footnote{The approach of \citet{Koehler2015} is to calculate grain optical properties for denser regions in the ISM without solving the time-dependent coagulation equation \citep{Ossenkopf1993, Ormel2009}.}, we assume that the dust properties change with increasing local density through accretion and coagulation. First, in the transition between the diffuse ISM and dense clouds, a second mantle can form on the surface of the CM grains, due to the coagulation of the small aromatic-rich carbon grains on top of the bigger grains, which might be  subsequently hydrogenated (a-C $\rightarrow$ a-C:H) and/or to the accretion of a-C:H material from the gas phase. Such carbonaceous mantles are efficiently processed by UV photons \citep{Alata2014}, but can stay H-rich as long as the radiation field is attenuated \citep{Faraday2014} or if the rehydrogenation process is efficient enough. This leads to grains with two mantles (core/mantle/mantle grains, CMM). Second, inside dense clouds, the CMM grains can coagulate into aggregates (AMM). On average, regarding the material abundance, one aggregate is composed of three CMM grains with amorphous silicate cores and one CMM grain with an amorphous carbon core. The formation of ice mantles on the surface of the aggregates (AMMI) can also occur in the densest regions, where the shielding from energetic photons is efficient enough to allow either gas molecules to form and to freeze out on the grains or surface chemistry to proceed effectively. The three types of evolved grains (CMM, AMM, and AMMI) contain 406~ppm of C in agreement with \citet{Parvathi2012}, who found that $355 \pm 64$~ppm of C are enclosed within grains for lines-of-sights where $N_H \gtrsim 2 \times$~$10^{21}$~H/cm$^2$ (equivalently where $E(B-V) \gtrsim 0.4$ or $A_{\rm V} \gtrsim 2$). For each grain type, a size distribution consistent with grain growth is considered as explained in \citet{Koehler2015} and shown in Fig.~\ref{Fig2}. For the CMM grains, $\sim 65$\% of the dust mass is in grains with $40~{\rm nm} \leqslant a \leqslant 0.25$~$\mu$m, $\sim 25$\% in grains with $0.25 \leqslant a \leqslant 0.5$~$\mu$m, and $\sim 10$\% in grains with $0.5 \leqslant a \leqslant 0.7$~$\mu$m. For the AMM (AMMI) aggregates, $\sim 50$\% of the dust mass is in grains with $48~(91)~{\rm nm} \leqslant a \leqslant 0.25$~$\mu$m, $\sim 40$\% in grains with $0.25 \leqslant a \leqslant 0.5$~$\mu$m, and $\sim 10$\% in grains with $0.5 \leqslant a \leqslant 0.7$~$\mu$m. The optical properties of all grains and a detailed description of the calculation method can also be found in \citet{Koehler2015}, while a more specific description of the grain scattering properties and efficiencies are given in paper I. A schematic view of the dust composition and stratification from the diffuse ISM to dense molecular clouds is shown in Fig.~\ref{Fig1}.

The models presented in Sects.~\ref{coreshine_cloudshine} and \ref{comparison_with_observations} take into account several dust population mixtures, which are labelled according to \citet{Koehler2015}:\\
$-$ CM model: no evolution, the cloud is uniformly filled by CM grains ;\\
$-$ CMM model: the formation of a second H-rich carbon mantle occured, the CM grains are replaced by CMM grains ;\\
$-$ CMM+AMM model: coagulation occured as well, the outer layers of the cloud are filled with CMM grains, while AMM grains are present in the layers in which the density is higher than a given density threshold, $\rho_t$ ;\\
$-$ CMM+AMMI model: same as the previous case but including the formation of ice mantles on the aggregates.\\
In the following, unless otherwise stated, $\rho_t$ is $1\,500$~H/cm$^3$ according to \citet[][see their Tab.~3]{Ysard2013}. These authors found that in the dense filament L1506 of the Taurus molecular complex, aggregates must prevail above such gas density threshold. Then, when a diffuse envelope surrounding the dense molecular cloud is assumed, it is always filled with CM grains only.

\subsection{Radiative transfer modelling: emission and scattering}
\label{modelling_tools}

\begin{figure}[!t]
\centerline{
\includegraphics[width=0.4\textwidth]{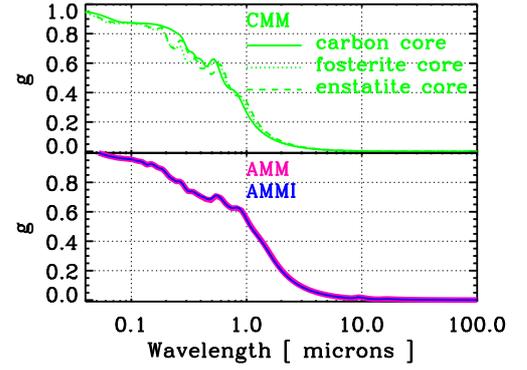}}
\caption{Scattering asymmetry factors $g$ for the dust particles presented in Sect.~\ref{dust_model}. The anisotropy factors are plotted for the size matching the peak of the mass distribution presented in Fig.~1 from \citet{Koehler2015}. {\it Top:} CMM particles with an amorphous carbon core (solid line), a core with the normative composition of enstatite (dashed line), and of forsterite (dotted line). {\it Bottom:} aggregate particle with (blue line) and without (magenta line) an ice mantle.}
\label{Fig3} 
\end{figure}

The radiative transfer calculations are performed with the coupling of the Monte-Carlo radiative transfer code CRT \citep{Juvela2005} and the dust emission and extinction code DustEM \citep{Compiegne2011}, as described in \citet{Ysard2012}. For the scattering, we use the Henyey-Greenstein phase function approximation \citep{HG1941}. The anisotropy factors of our various dust particles are plotted in Fig.~\ref{Fig3}. This figure shows that the smallest particles scatter light almost isotropically ($g \sim 0$) in the wavelength range relevant for this paper (near-IR and mid-IR). For the largest particles, the scattering is still close to isotropy in the mid-IR and $g$ remains below 0.3 in the near-IR. Since all CMM, AMM, and AMMI grains are small compared to the wavelengths of interest, we assume that the Henyey-Greenstein phase function is a good approximation for our study.

\begin{figure}[!t]
\centerline{\includegraphics[width=0.4\textwidth]{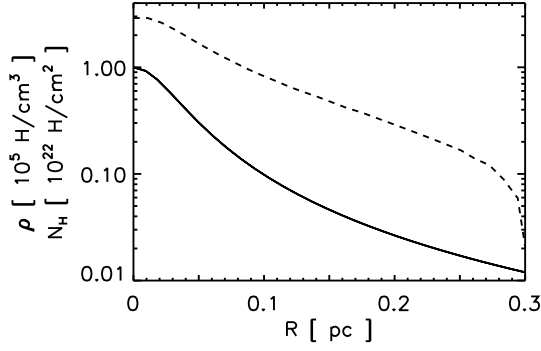}}
\caption{Density distribution (solid line) and colum density profile (dashed line) for Eq.~\ref{eq_density_distribution} with $\rho_C = 10^5$~H/cm$^3$, $\rho_{{\rm DIM}} = 0$~H/cm$^3$, $R_{out} = 0.3$~pc, $R_{flat} = 0.03$~pc, and $p = 2$ (see Sect.~\ref{coreshine_cloudshine}).}
\label{Fig4} 
\end{figure}

Analysis of far-IR observations of dense molecular regions have shown that for these clouds the gas density distribution is usually well approximated by a Plummer-like function \citep[see for instance][]{Arzoumanian2011, Juvela2012}:
\begin{eqnarray}
\label{eq_density_distribution}
\rho(r) &=& \frac{\rho_C}{\left[ 1 + \left(r/R_{flat}\right)^2\right]^{p/2}} \; {\rm for} \; r \leqslant R_{out},\\
&=& \rho_{{\rm DIM}} \;\;\;\;\;\;\;\;\;\;\;\;\;\;\;\;\;\;\;\;\, {\rm for} \; R_{out} < r \leqslant 0.5 \; {\rm pc,}\nonumber
\end{eqnarray}
where $\rho_C$ is the central hydrogen density, $\rho_{{\rm DIM}}$ is the density of a possible diffuse envelope surrounding the cloud, $R_{flat}$ is the radius characterising the density profile width, $p$ is a positive constant characterising the density profile steepness, and $R_{out}$ is the external radius of the cloud. In the following, we always assume spherical clouds and gas density distributions following Eq.\ref{eq_density_distribution} (see Fig.~\ref{Fig4}). We perform full 3D calculations by discretizing the cloud cells on a cartesian grid.

\begin{figure}[!t]
\centerline{\includegraphics[width=0.4\textwidth]{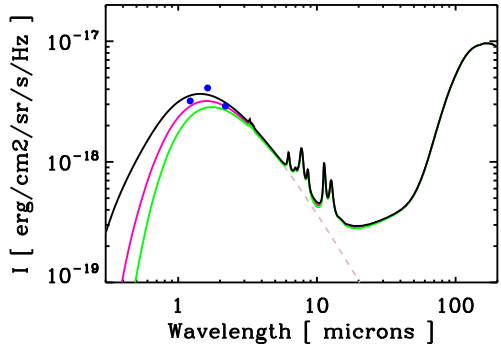}}
\caption{The black line shows the ISRF+CM radiation field intensity (the grey dashed line shows the ISRF alone). The pink and the green lines show the same but extinguished by a layer of CM grains with $A_{{\rm V}}^{ext} = 1$ and 2$^{{\rm mag}}$, respectively. The blue circles show the near-IR radiation field estimated by \citet{Lehtinen1996} for a Galactic latitude of about $\pm 20\degr$.}
\label{Fig5} 
\end{figure}

The last parameter to determine is the radiation field illuminating the cloud. From the UV to the near-IR, the radiation field is dominated by starlight and for the clouds to which our model is compared in this paper (Sect. \ref{comparison_with_observations}), the standard interstellar radiation field (ISRF) as defined by \citet{Mathis1983} at a galactocentric distance of 10~kpc is assumed to be a good approximation. However, as our aim is to study the dust scattering in the mid-IR, ISRF alone is not enough and a dust emission component has to be added. In the following, 21 starless cores are considered to compare to the model calculations, 13 of which are located in the Taurus-Perseus molecular complex \citep{Lefevre2014}. In this region, the dense clouds exhibiting mid-IR scattering are also detected in $^{13}$CO and embedded in larger clouds with lower densities and detected in $^{12}$CO \citep{Narayanan2008, Qian2012, Meng2013}. These more diffuse envelopes have visual extinctions $A_{\rm V} \sim 1$ \citep{Padoan2002} and IR SEDs consistent with diffuse ISM-type dust \citep{Flagey2009}. Consequently, we assume that the mid-IR photons emitted by these surrounding envelopes are well described by the emission of CM dust, with a column density $N_{\rm H} \sim 10^{21}$~H/cm$^2$, heated by the ISRF. The resulting radiation field is displayed in Fig.~\ref{Fig5} and referred to as the ISRF+CM radiation field. The other starless cores considered in Sect.~\ref{comparison_with_observations} are located in L183, Chamaeleon (4 cores), and Cepheus (3 cores), for which we assume the ISRF+CM radiation field to be a good approximation \citep[see for instance][for the Chamaeleon]{AlvesDeOliveira2014}. This description of the radiation field as well as the description of the cloud density distribution, even if simplistic, seem reasonable as our aim in this study is to demonstrate the ability of our model to account for the observed mid-IR emission.

\section{THEMIS scattering spectrum}
\label{coreshine_cloudshine}

\begin{figure}[!t]
\centerline{\includegraphics[width=0.4\textwidth]{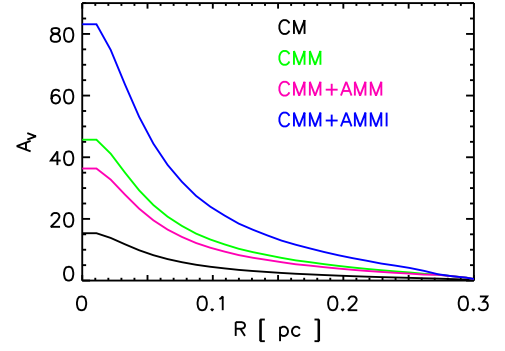}}
\caption{Extinction profiles in the V photometric band ($\lambda = 0.55~\mu$m) for the control cloud described in Sect.~\ref{coreshine_cloudshine} and the dust population mixtures presented in Sect.~\ref{dust_model}: CM in black, CMM in green, CMM+AMM in magenta, and CMM+AMMI in blue.}
\label{Fig6} 
\end{figure}

\begin{figure}[!t]
\centerline{\includegraphics[width=0.4\textwidth]{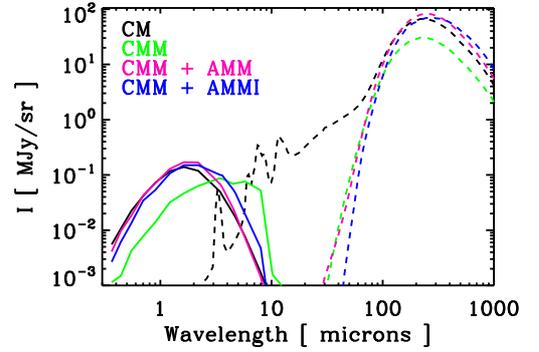}}
\caption{Scattering (solid lines in the near- to mid-IR) and emission (dashed lines in the mid- to far-IR) spectra of the dust populations described in Sect.~\ref{dust_model} at the centre of the control cloud defined in Sect.~\ref{coreshine_cloudshine}. The black lines show the case of CM grains, the green lines of CMM grains, the magenta lines of CMM+AMM grains, and the blue lines of CMM+AMMI grains.}
\label{Fig7} 
\end{figure}

\begin{figure*}
\centerline{
\begin{tabular}{c}
\includegraphics[angle=90,width=14cm,height=5cm]{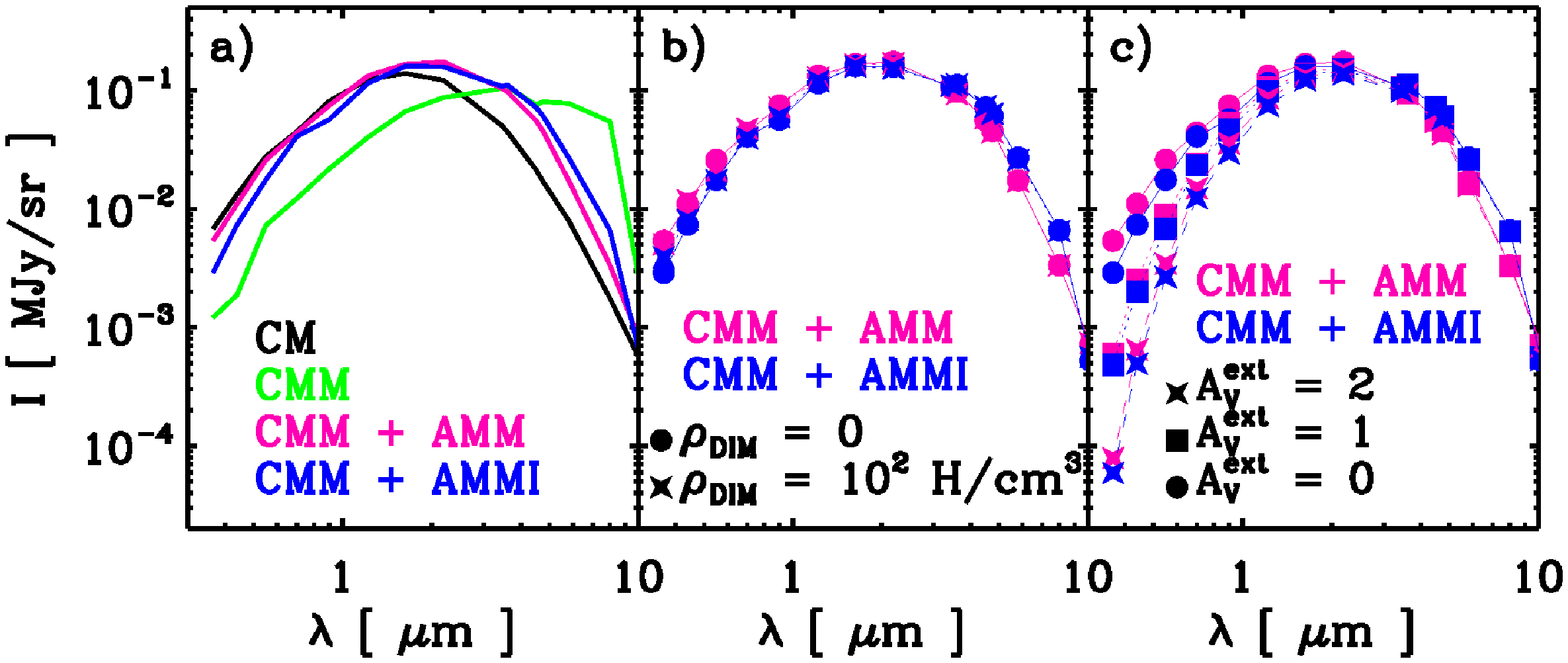} \\
\includegraphics[angle=90,width=14cm,height=5cm]{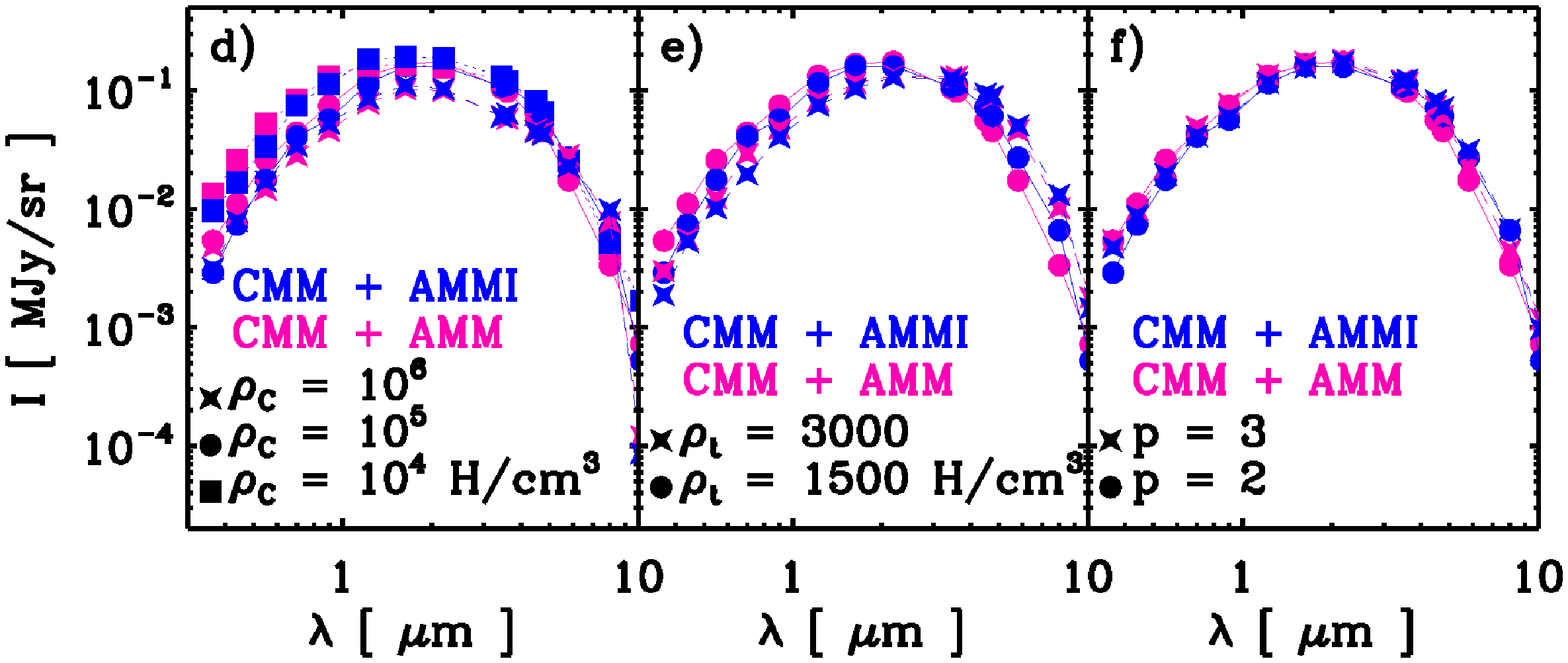} 
\end{tabular}}
\caption{{\it (a)} Scattering spectra of the dust populations described in Sect.~\ref{dust_model} at the centre of the control cloud defined in Sect.~\ref{coreshine_cloudshine}. {\it (b)} Influence of a surrounding diffuse envelope on the scattering spectrum. {\it (c)} Influence of the radiation field intensity. {\it (d)} Influence of the cloud central density. {\it (e)} Influence of the density at which aggregates are the dominant dust population. {\it (f)} Influence of the cloud density profile steepness.}
\label{Fig8}
\end{figure*}

\begin{figure*}
\centerline{
\begin{tabular}{c}
\includegraphics[angle=90,width=14cm,height=5cm]{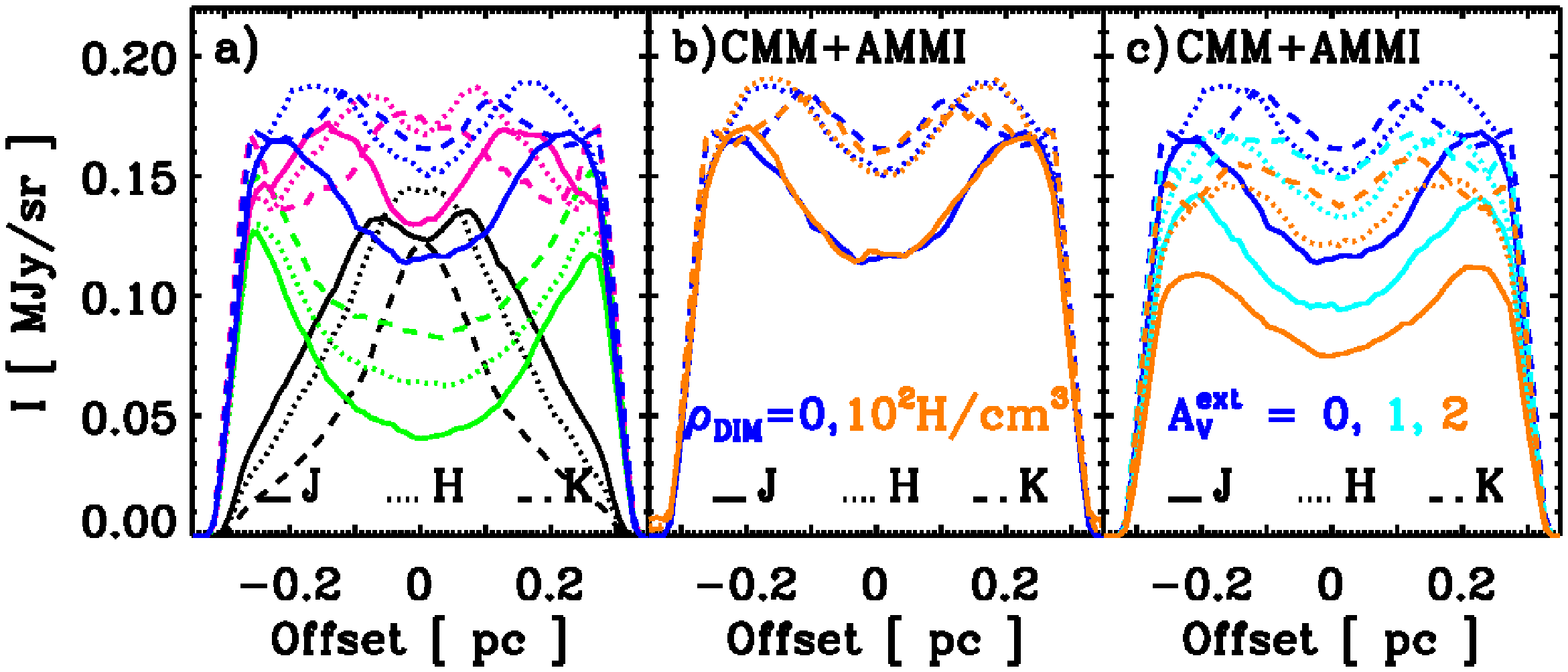} \\
\includegraphics[angle=90,width=14cm,height=5cm]{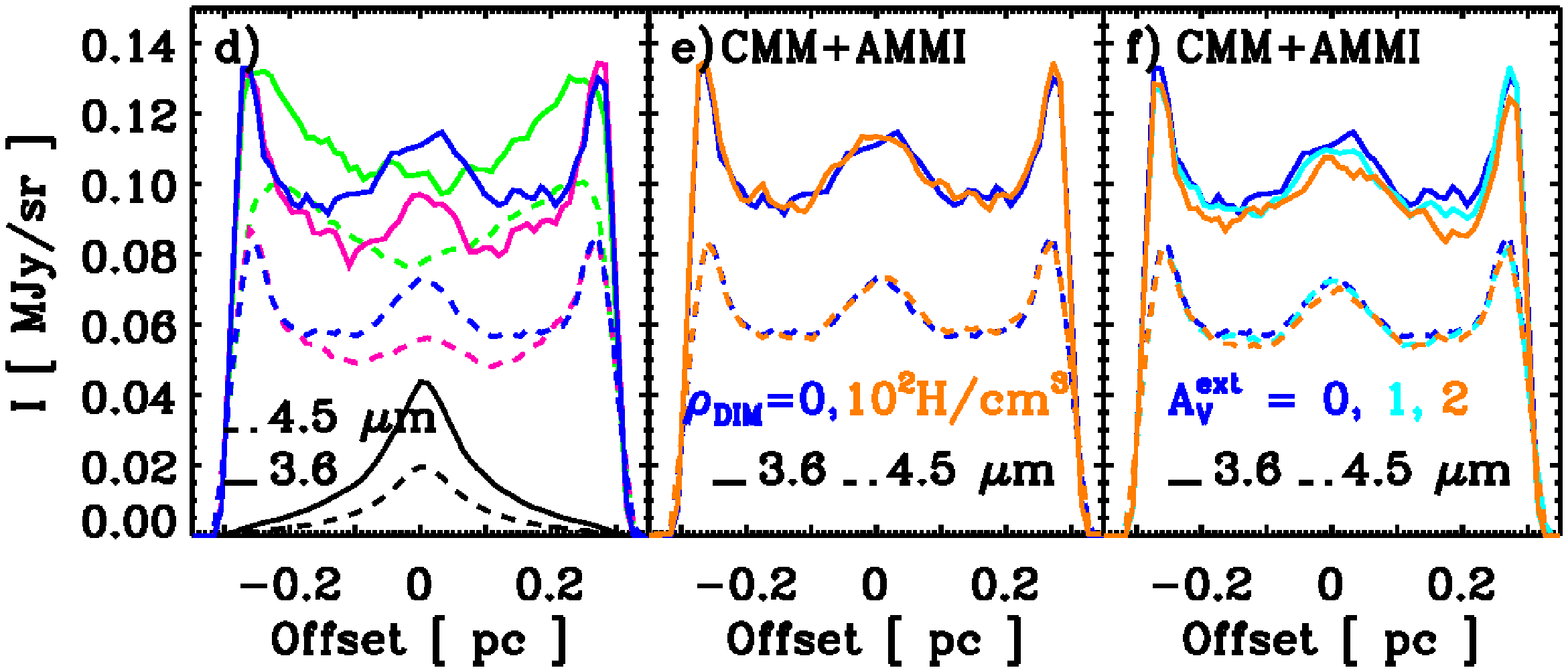} \\
\includegraphics[angle=90,width=14cm,height=5cm]{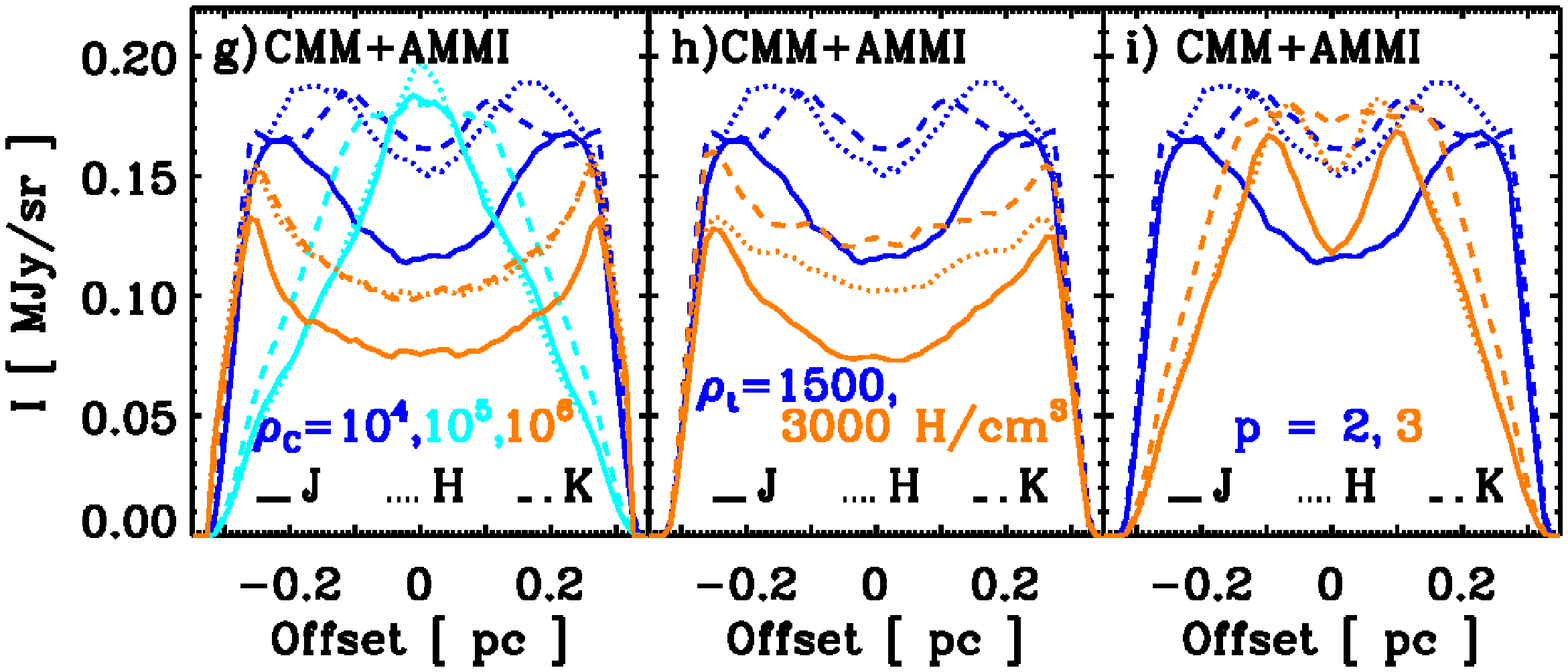} \\
\includegraphics[angle=90,width=14cm,height=5cm]{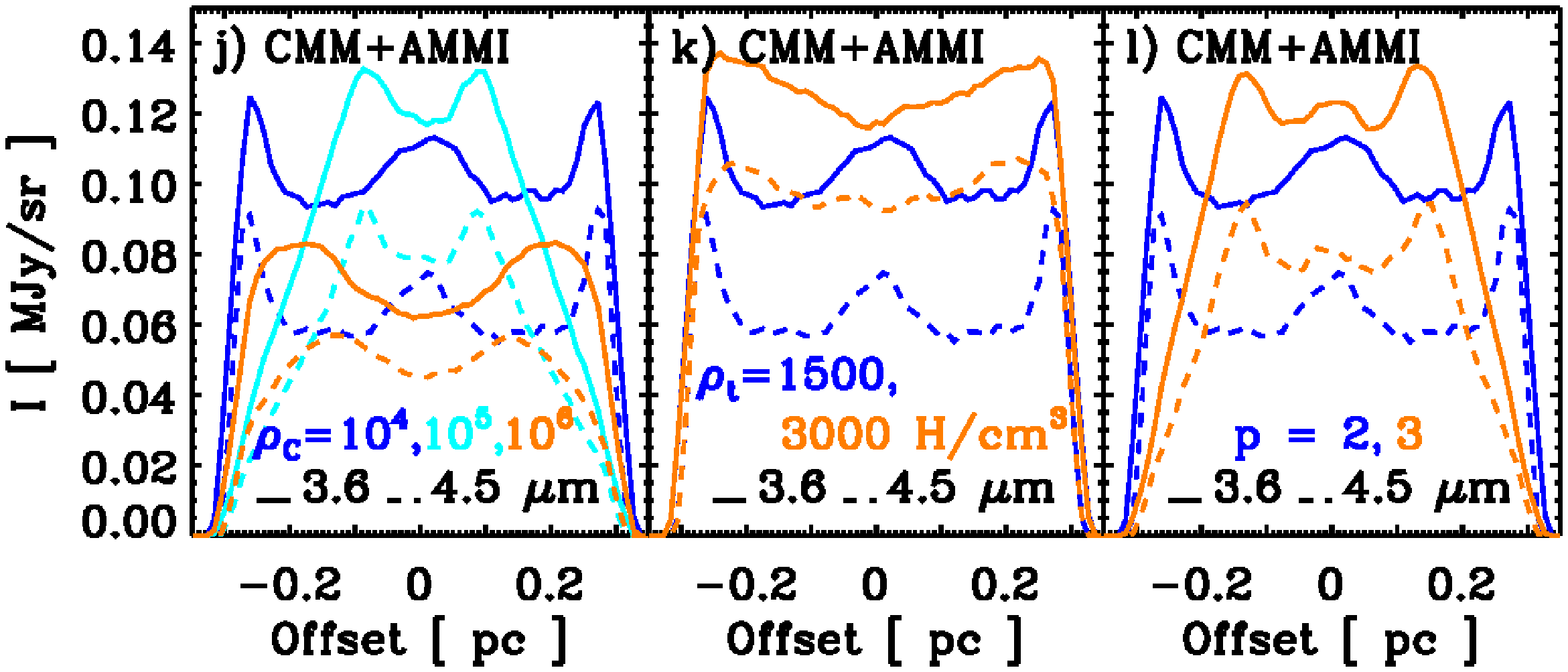}
\end{tabular}}
\caption{{\it (a, d)} Near-IR (J, H, and K photometric bands) and mid-IR (3.6 and 4.5~$\mu$m) radial profiles of the dust populations described in Sect.~\ref{dust_model} for the control cloud defined in Sect.~\ref{coreshine_cloudshine}. {\it (b, e)} Influence of a surrounding diffuse envelope on the scattering spectrum. {\it (c, f)} Influence of the radiation field intensity. {\it (g, j)} Influence of the cloud central density. {\it (h, k)} Influence of the density at which aggregates are the dominant dust population. {\it (i, l)} Influence of the cloud density profile steepness.}
\label{Fig9}
\end{figure*}

We investigate the influence of the various parameters defining the level of scattered light coming from dense clouds. Our starting point is a spherical cloud with $\rho_C = 10^5$~H/cm$^3$, $\rho_{DIM} = 0$, and according to the observational result of \citet{Arzoumanian2011}, which is that the column density distribution FWHM of dense clouds is on average of 0.1~pc, we assume $R_{out} = 0.3$~pc, $R_{flat} = 0.03$~pc, and $p = 2$ (see Eq.~\ref{eq_density_distribution} and Fig.~\ref{Fig4}). These parameters lead to a cloud with a central column density of $2.9 \times 10^{22}$~H/cm$^2$ and an H mass of $8 M_{\odot}$ (with $0.6 M_{\odot}$ enclosed within the FWHM/2 radius), which is then illuminated by the ISRF+CM radiation field (see Sect.~\ref{modelling_tools} and Fig.~\ref{Fig5}) and populated with the dust population mixtures presented in Sect.~\ref{dust_model}: CM, CMM, CMM+AMM, and CMM+AMMI, with $\rho_t = 1\,500$~H/cm$^3$ at an offset of 0.24~pc from the cloud centre. The corresponding cloud dust masses enclosed in the FWHM/2 radius are $3.4 \times 10^{-3} M_{\odot}$ for CM, $6 \times 10^{-3} M_{\odot}$ for CMM and CMM+AMM, and $1.2 \times 10^{-2} M_{\odot}$ for CMM+AMMI. In the following, this set of parameters is referred to as the control cloud. The corresponding extinction profiles are shown in Fig.~\ref{Fig6} and the results are presented in Figs.~\ref{Fig7}, \ref{Fig8} and \ref{Fig9}.

The emission and scattering spectra at the centre of the control cloud are shown in Fig.~\ref{Fig7}. As expected, the CM grains (black lines), even if efficient scatterers in the near- and mid-IR, exhibit strong emission in the IRAC 5.8 and 8~$\mu$m bands and have thus to be ruled out from the plausible C-shine explanations. The CMM grains (green lines) have also to be discarded as light scattering is as efficient in the IRAC 3.6 and 4.5~$\mu$m bands as in the 5.8 and 8~$\mu$m bands (Fig.~\ref{Fig8}a). As explained in paper I, this enhancement comes from the addition of the H-rich carbon mantle (see their Fig. 10). On the contrary, the CMM+AMM and CMM+AMMI populations seem to be good candidates to explain the C-shine observations (magenta and blue lines, respectively in Fig.~\ref{Fig8}a and \ref{Fig9}a). Indeed, the aggregates appear to scatter light efficiently only up to $\sim 5~\mu$m, thus matching the coreshine requirement of being detected in emission in the first two IRAC bands (3.6 and 4.5~$\mu$m) and in extinction in the two other IRAC bands (5.8 and 8~$\mu$m). The contribution of the largest aggregates ($0.5 \leqslant a \leqslant 0.7~\mu$m, which make up 10\% of the dust mass) to the scattered intensity is less than 5\% in the J and H photometric bands, $\sim 10$\% in the K band, $\sim 25$\% at 3.6~$\mu$m, and $\sim 30$\% at 4.5~$\mu$m. For both aggregate types, the scattering spectrum peak is located in the near-IR, where the cloudshine is observed. However, the shape and strength of their near- and mid-IR radial profiles differ (Fig.~\ref{Fig9}a and d) and should help to discriminate between the two dust population mixtures: CMM+AMM or CMM+AMMI. Fig.~\ref{Fig9} further shows that the near- and mid-IR profiles do not follow the same trend at the cloud centre. Indeed, in the near-IR, where absorption by dust grains is efficient, the profiles mostly reflect the number of available photons to scatter. In the mid-IR, where absorption is less efficient, this effect is mitigated by column density variations. In particular, in Figs.~\ref{Fig9}d, e, f, j, k, and l, the increase seen at the centre of the mid-IR profiles matches the radius characterising the density profile width, $R_{flat}$ (see Eq.~\ref{eq_density_distribution}).

There are several ways of modifying the density distribution: changing $\rho_C$ (Fig.~\ref{Fig8}c and Figs.~\ref{Fig9}i and l, $\rho_C = 10^4, 10^5$, and $10^6$~H/cm$^3$), $p$ (Fig.~\ref{Fig8}f and Figs.~\ref{Fig9}k and n, $p = 2$ and 3), and $\rho_t$ (Fig.~\ref{Fig8}d and Figs.~\ref{Fig9}j and m, $\rho_t = 1\,500$ and $3\,000$~H/cm$^3$). Increasing (decreasing) the central density leads to a significant decrease (increase) in the scattering spectrum. This is due to more (less) efficient absorption of the ISRF+CM photons by the outer layers of the cloud for the near-IR part. Absorption in the mid-IR is less efficient and, as a result, the decrease in central regions is small for $\rho_C = 10^4$ to $10^5$~H/cm$^3$ but is significant when reaching $10^6$~H/cm$^3$. Colour ratios at C-shine wavelengths are thus strongly dependent on the cloud central density. The radial profile shapes also vary strongly from profiles increasing by a factor of $\sim 3$ from the edge to the cloud centre for $\rho_C = 10^4$~H/cm$^3$, to profiles decreasing by a factor of $\sim 2$ for $\rho_C = 10^6$~H/cm$^3$. Then, changing the $p$-factor does not produce strong variations in the cloud central scattering spectrum (Fig.~\ref{Fig8}f). However, its influence on the radial profiles is quite significant (Figs.~\ref{Fig9}i and l). In the case of the control cloud, increasing $p$ means decreasing the density at the edge of the cloud by a factor $\sim 10$ and keeping the same density for $r \lesssim 0.01$~pc. This results in lower scattering at the edge of the cloud because of lower column density but to higher scattering at the centre due to less efficient IR photon absorption at the edge. Finally, increasing the threshold density at which the aggregates appear from $\rho_t = 1\,500$~H/cm$^3$ in the control cloud to 3\,000~H/cm$^3$, changes the scattering spectrum with a decrease in the near-IR and an increase in the mid-IR (Figs.~\ref{Fig8}e and Figs.~\ref{Fig9}h and k). The higher proportion of CMM grains in the total dust column density compared to AMM(I) grains explains these changes: CMM grains absorb more in the near-IR and are more efficient scatterers at longer wavelengths than AMM(I) grains (see Fig.~\ref{Fig7} and the optical properties presented in paper I).

Dense clouds are usually embedded in larger and more diffuse structures. To extract the C-shine signal, two ways of dealing with observations are then possible. Either the diffuse envelope is subtracted from the core emission but then the ISRF+CM radiation field heating the dense core may be partly extinguished. Or the diffuse envelope is not removed and its contribution to the emission/scattering has to be taken into account. To illustrate the first option, we consider the control cloud but illuminated by the ISRF+CM radiation field extinguished by a layer of CM grains with $A_{{\rm V}}^{ext} = 1$ and 2 (Fig.~\ref{Fig5}). The consequences are a global decrease in the near-IR scattering and no changes in the mid-IR (Fig.~\ref{Fig8}c and Figs.~\ref{Fig9}c and f). For the second option, we consider that the control cloud is surrounded by a diffuse envelope with a density $\rho_{DIM} = 100$~H/cm$^3$ for $R_{out} \leqslant r \leqslant 0.5$~pc. The effect is small with the envelope accounting only for an extra 0.01~MJy/sr in the near-IR and negligible in the mid-IR (Figs.~\ref{Fig8}b and Figs.~\ref{Fig9}b and e).

\section{Comparison with C-shine observations}
\label{comparison_with_observations}

Within the framework of our global dust modelling approach THEMIS, presented in the previous sections, our aim is now to explain the observed near- and mid-IR trends in the dust SED towards dense molecular clouds. Following the evolutionary path described in Fig.~\ref{Fig1}, we demonstrate the ability of THEMIS to reproduce the C-shine observations in the mid-IR (Sect.~\ref{coreshine_modelling}) and in the near-IR (Sect.~\ref{cloudshine_modelling}).

\subsection{Coreshine}
\label{coreshine_modelling}

\begin{figure*}[!t]
\centerline{
\begin{tabular}{cc}
\includegraphics[width=0.37\textwidth]{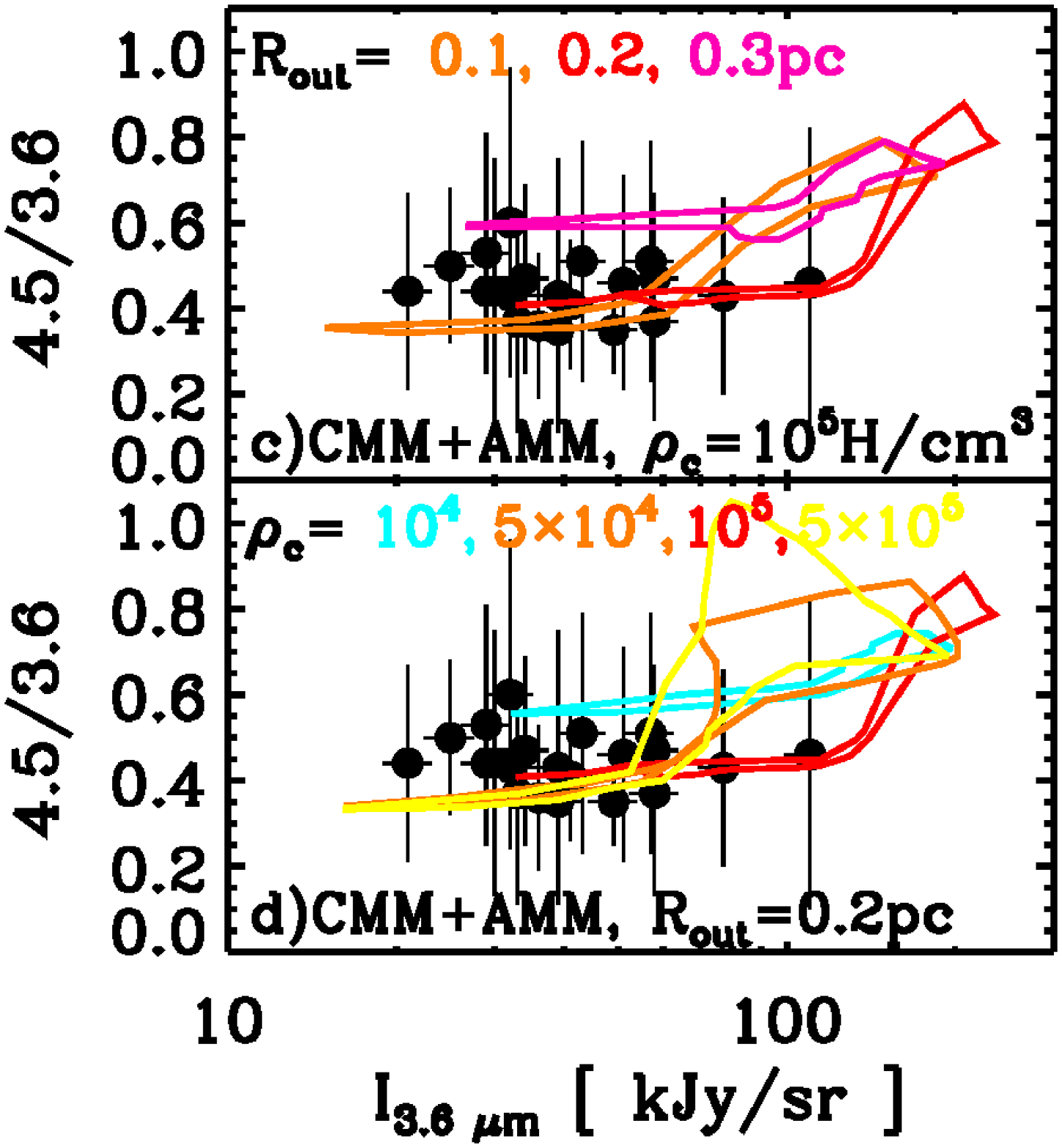} & \includegraphics[width=0.37\textwidth]{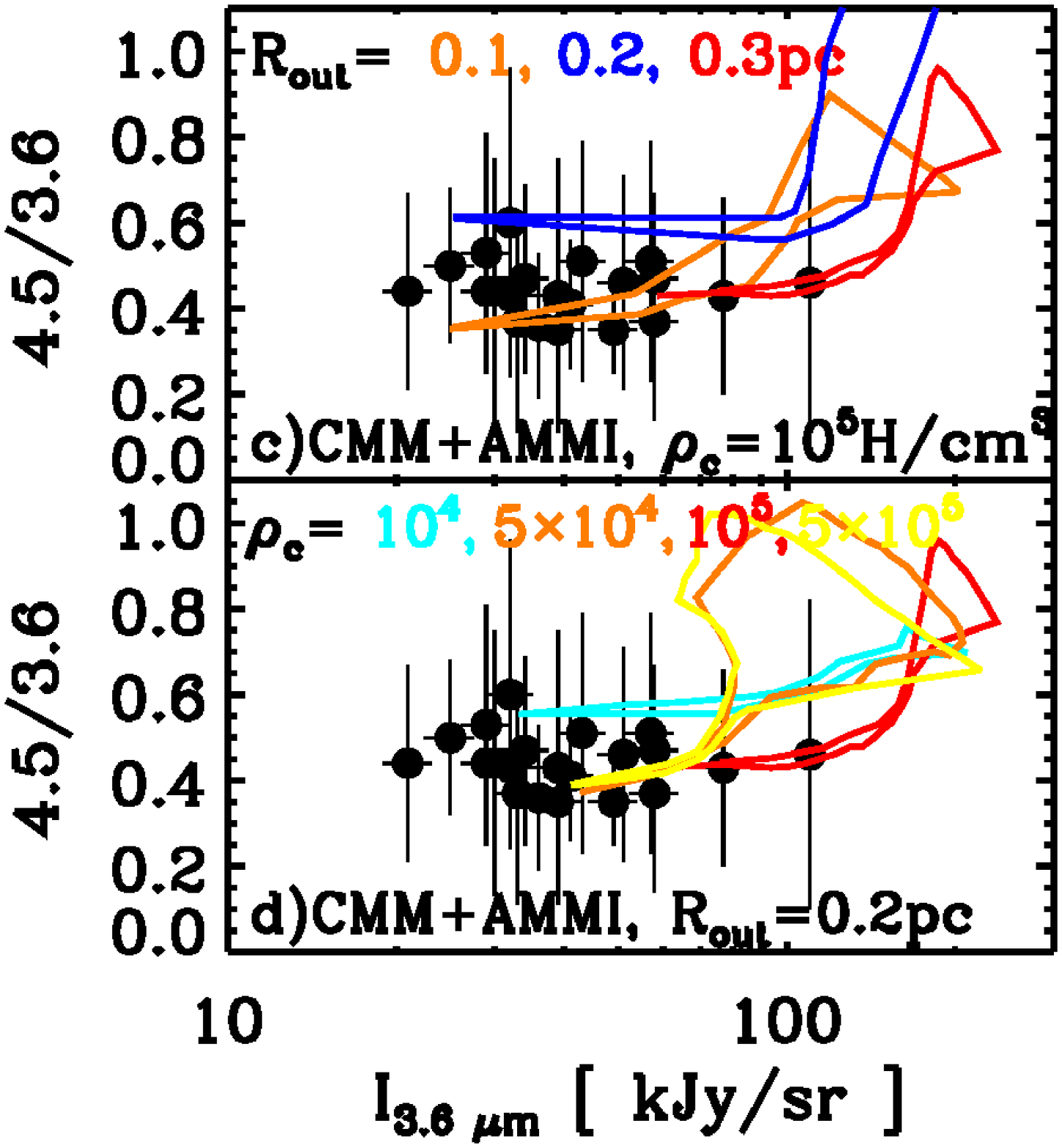} \\
\includegraphics[width=0.35\textwidth]{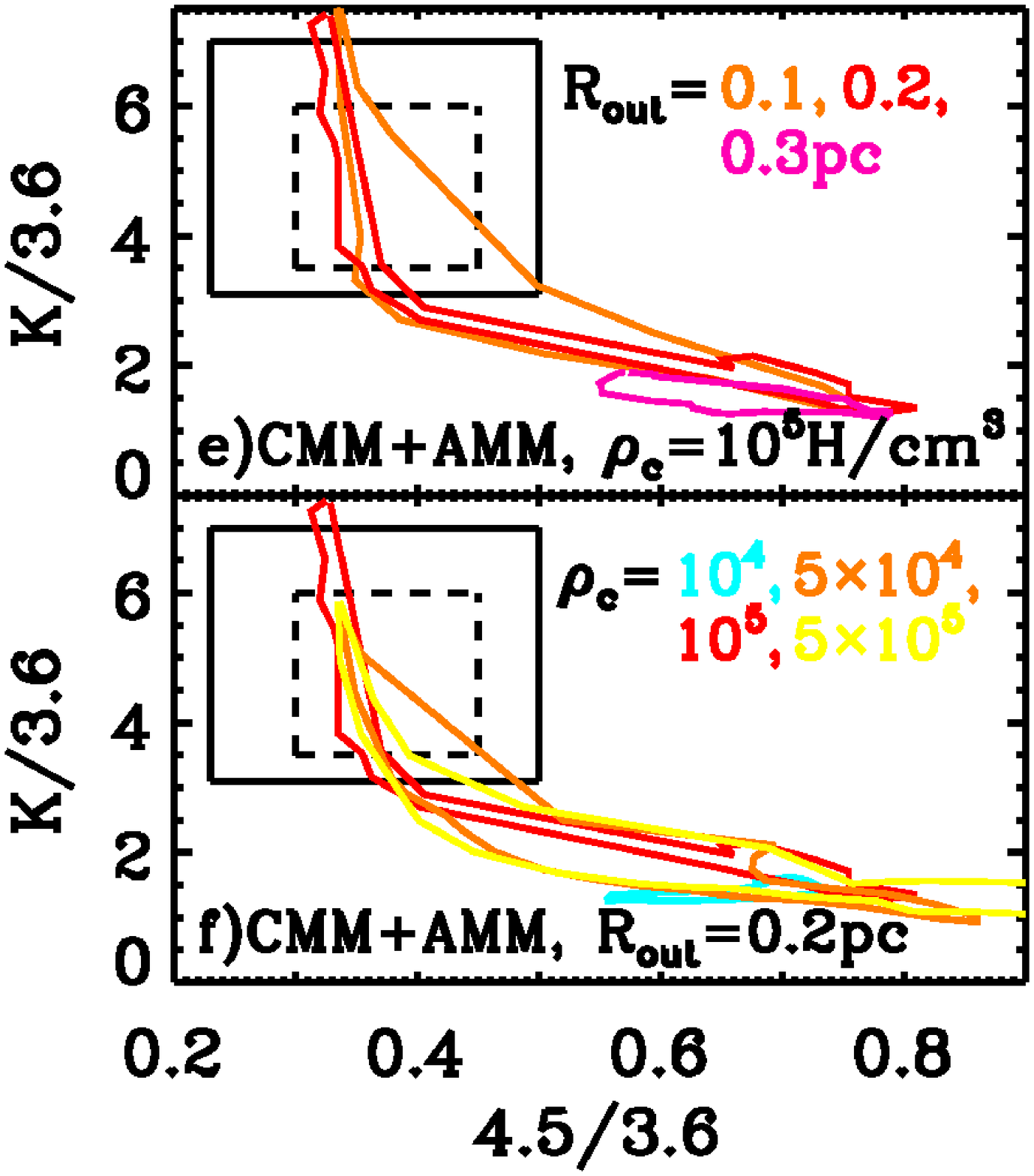} & \includegraphics[width=0.35\textwidth]{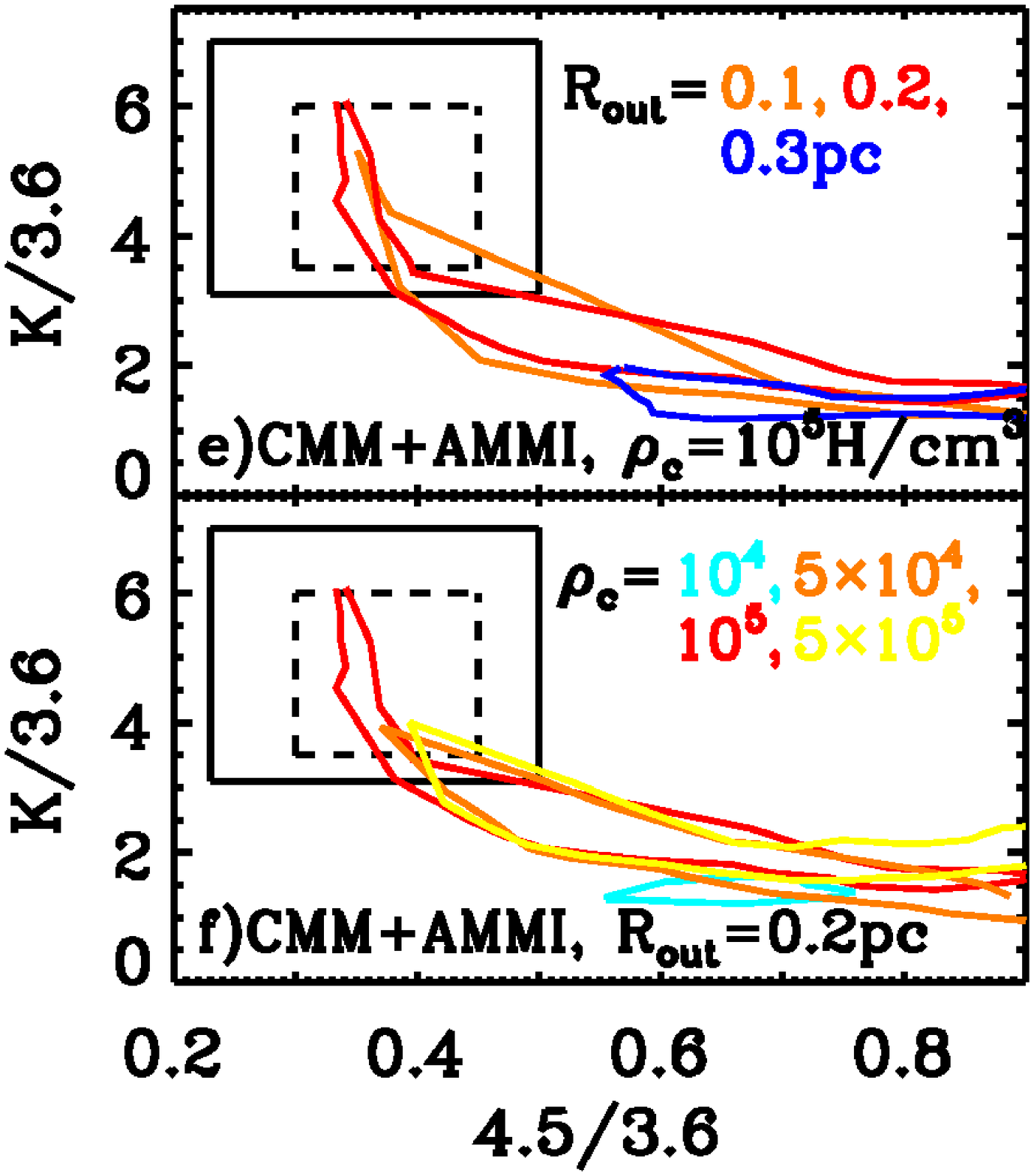}
\end{tabular}}
\caption{The areas enclosed in coloured lines show the spread of the results for model clouds filled with CMM+AMM grains ({\it a}, {\it b}, {\it e}, and {\it f}) and with CMM+AMMI grains ({\it c}, {\it d}, {\it g}, and {\it h}). The spread is representative of both variations along the sphere radius and variations in $f_{back}$ from 0 to 1 (from left to right in all figures, see Sect.~\ref{coreshine_modelling} and footnote~\ref{fback}). {\it (a)}, {\it (b)}, {\it (c)}, and {\it (d)} 4.5 to 3.6~$\mu$m coreshine ratio as a function of the 3.6~$\mu$m intensity. The black circles are the coreshine ratios measured by \citet{Lefevre2014} towards 21 starless cores and the black lines the scatter in these measurements. {\it (e)}, {\it (f)}, {\it (g)}, and {\it (h)} Near- to mid-IR ratio as a function of the 4.5 to 3.6~$\mu$m coreshine ratio. The solid and dashed boxes show the range of values measured towards the Taurus-Perseus regions and L183/L134 cloud, respectively (Lef\`evre et al. 2014, see their Figs. 17 \& 18). In Figs.~{\it a} and {\it e} ({\it c} and {\it g}), the magenta (blue) lines show the results for the control cloud parameters defined in Sect.~\ref{coreshine_cloudshine} ($R_{out} = 0.3$~pc, $\rho_C = 10^5$~H/cm$^3$) and the orange and red lines for the same set of parameters except that $R_{out} = 0.1$ and 0.2~pc, respectively. In Figs.~{\it b} and {\it f} ({\it d} and {\it h}), the red lines show the same as in the previous figures ($R_{out} = 0.2$~pc, $\rho_C = 10^5$~H/cm$^3$), the cyan lines show the results for $\rho_C = 10^4$~H/cm$^3$, the orange lines for $\rho_C = 5 \times 10^4$~H/cm$^3$, and the yellow lines for $\rho_C = 5 \times 10^5$~H/cm$^3$.}
\label{Fig10} 
\end{figure*}

All coreshine observations were obtained with the IRAC instrument onboard the Spitzer observatory and are gathered in \citet{Paladini2014} and \citet{Lefevre2014}, from which we selected 21 starless cores in the Taurus-Perseus, Chamaeleon, Cepheus, and L183/L134 regions (see their Fig. 9 and Tab. 1). \citet{Lefevre2014} summarised their results in two figures that we reproduce here: the 4.5 to 3.6~$\mu$m ratio, that they name ``coreshine ratio'', as a function of the 3.6~$\mu$m intensity and the 2.2 to 3.6~$\mu$m ratio, that they name the ``near-IR to mid-IR ratio'', as a function of the coreshine ratio. For the model cloud, we use the parameters of the control cloud defined in Sect.~\ref{coreshine_cloudshine}. After convolving our models with a $10\arcsec$ FWHM Gaussian kernel to simulate the data analysis presented in \citet{Lefevre2014} and following the \citet{Lehtinen1996} prescription to take into account the part of the ISRF+CM light that can be transmitted through the cloud\footnote{\label{fback}The CM part of the ISRF+CM radiation field represents the photons emitted in the direct surroundings of the cloud. The fraction of this emission that comes from the background of the cloud, $f_{back}$, can be both scattered or transmitted with $I_{transmitted} = f_{back} {\rm CM} \times e^{-\tau}$, where $\tau$ is the cloud opacity at given wavelength and radial position.}, we compute the synthetic photometry for each pixel along a radial cut through our model clouds. The results are shown in Figs.~\ref{Fig10}a and b, which present the coreshine ratio as a function of the 3.6~$\mu$m intensity for the 21 starless cores of \citet{Lefevre2014} and in Fig.~\ref{Fig10}c and d, which displays the near- to mid-IR ratio as a function of the coreshine ratio for the Taurus-Perseus and L183/L134 regions. The CMM model has to be ruled out to explain coreshine since it only marginally fits the coreshine ratio and fails at reproducing the near- to mid-IR ratio (green areas in Fig.~\ref{Fig10}). On the contrary, the CMM+AMM and CMM+AMMI models can explain the coreshine observations (magenta and blue areas in Fig.~\ref{Fig10}, respectively). Based on the results presented in Sect.~\ref{coreshine_cloudshine} and Fig.~\ref{Fig7}, the cloud parameters are as important as the dust model to explain the dispersion in the observations. Varying the cloud external radius from $R_{out} = 0.3$ to 0.1~pc (Figs.~\ref{Fig10}a and c) and the central density from $\rho_C = 10^4$ to $5 \times 10^5$~H/cm$^3$ (Figs.~\ref{Fig10}b and d) seem enough to explain most of the observed scatter.

\subsection{Cloudshine}
\label{cloudshine_modelling}

\begin{figure}[!t]
\centerline{\includegraphics[width=0.4\textwidth]{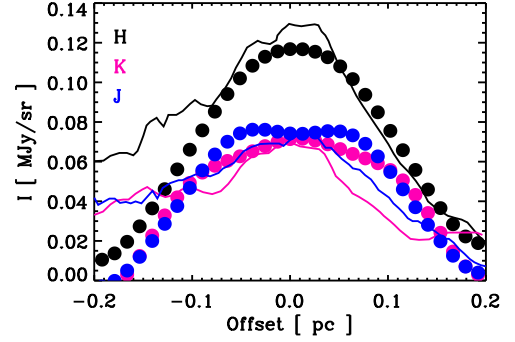}}
\caption{Cloudshine radial profiles across the TMC-1N interstellar filament. The blue, black, and red lines show the profiles measured by \citet{Malinen2013} in the J, H, and K photometric bands of the WFCAM instrument, respectively, smoothed to 40'' angular resolution. The circles of the same colours show our fit to these data (CMM+AMMI and $A_{{\rm V}}^{ext} = 1.5^{{\rm mag}}$, see Sect.~\ref{cloudshine_modelling} for details about the modelling). Our model is smoothed to the same angular resolution as the data and the synthetic photometry in the WFCAM filters is computed at each position.}
\label{Fig11}
\end{figure}

As described in Sect.~\ref{introduction}, cloudshine is observed in the near-IR, usually in the three photometric bands J, H, and K. For instance, using the Wide Field CAMera of the United Kingdom InfraRed Telescope (WFCAM of UKIRT), \citet{Malinen2013} observed a $1\degr \times 1\degr$ field in the Taurus Molecular Complex ($d \sim 140$~pc). This field is centred on the dense TMC-1N cloud (RA (J2000) 4h39m36s and Dec (J2000) $+26\degr 39' 22''$) north of TMC-1. After smoothing the data to $40\arcsec$ resolution on a $8\arcsec$ pixel grid and removing background emission, \citet{Malinen2013} were able to extract J, H, and K coreshine median radial profiles across the TMC-1N filament (see their Fig.~12). This filament was previously observed with PACS and SPIRE instruments onboard Herschel: using the colour temperature of the dust submm emission and assuming that the dust opacity varies as 0.1~cm$^2$/g $(\nu / 1000~{\rm GHz})^{\beta}$ with $\beta = 2$, \citet{Malinen2012} found that the density distribution of the cloud is well described by a Plummer-like function with $\rho_C = 4 \times 10^4$~cm$^{-3}$, $R_{flat} = 0.03$~pc, and $p = 3$ (see Eq.~\ref{eq_density_distribution}). To model this cloud, we start from this density distribution and the dust population mixtures presented in Sect.~\ref{coreshine_modelling}. The cloud is then illuminated by the ISRF+CM radiation field, extinguished or not by a layer of CM grains with $A_V^{ext} = 0.5$ to 3 (see Fig.~\ref{Fig5}). The resulting dust scattered emission maps\footnote{In the near-IR, the optical properties of THEMIS model dust populations make the transmitted light contribution always negligible.} are finally smoothed to the same resolution as the data presented by \citet{Malinen2013}. The best fit is achieved for the CMM+AMMI populations illuminated by the ISRF+CM radiation field with $A_{{\rm V}}^{ext} = 1.5^{{\rm mag}}$. The results are presented in Fig.~\ref{Fig11}. As \citet{Malinen2013} subtracted the cloud background, it is not surprising to find $A_{{\rm V}}^{ext} \neq 0$: this probably reflects the fact that the incident radiation field is extinguished by the removed envelope before reaching the dense cloud. The peak intensity and the profile shapes are well reproduced for the three bands even if for the J and K bands the modelled profiles are slightly broader than the observed ones.

\begin{figure}[!t]
\centerline{\includegraphics[width=0.4\textwidth]{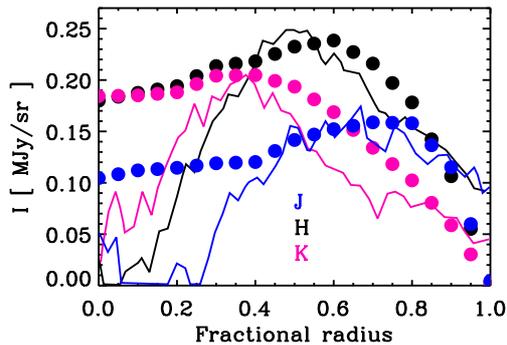}}
\caption{Cloudshine angle-averaged radial profiles across a dark core in L1451. The blue, black, and red lines show the profiles measured by \citet{Foster2006} in the OMEGA 2000 camera J, H, and K photometric bands, respectively. The circles of the same colours show our fit to these data (CMM+AMMI and radiation field from \citet{Lehtinen1996}, see Sect.~\ref{cloudshine_modelling} for details about the modelling). The synthetic photometry in the OMEGA 2000 filters is computed at each position.}
\label{Fig12}
\end{figure}

A second example of cloudshine observations in the J, H, and K photometric bands can be found in \citet{Foster2006}. Using the OMEGA 2000 camera on Calar Alto Observatory \citep[PSF $\sim 1\arcsec$, see][]{Kovacs2004}, these authors observed the B5, L1448, and L1451 regions in the Perseus molecular complex. After selecting a circular dark core in L1451 and assuming it to be spherical, \citet{Foster2006} produced angle-averaged radial brightness profiles in the J, H, and K bands. These profiles are shown in Fig.~\ref{Fig12} and cannot be reproduced whith a model cloud illuminated by the ISRF+CM radiation field. This produces H/J and H/K ratios too low by 10 to 20\% at the peak position, no matter what density distribution or dust population mixture are chosen. This comes from the high sensitivity of dust near-IR scattering to the local conditions and points the limit of our choice of an ``average'' radiation field when looking at high resolution data. To solve this problem, we use the results presented in \citet{Lehtinen1996}. Starting from DIRBE data, \citet{Lehtinen1996} estimated the diffuse emission in the J, H, and K bands as a function of Galactic latitude. At the latitude of L1451 ($b \sim -22\degr$), the brightnesses are 0.32, 0.41, and 0.29~MJy/sr in the J, H, and K bands, respectively, corresponding to -10, +12, and -5\% difference with respect to the ISRF+CM radiation field (blue circles in Fig.~\ref{Fig5}). The results shown in Fig.~\ref{Fig12} are obtained for a cloud with $\rho_C = 10^5$~H/cm$^3$, $R_{flat} = 0.03$~pc, $p = 3$, $\rho_t = 1\,500$~H/cm$^3$, $\rho_{DIM} = 100$~H/cm$^3$, and CMM+AMMI grains. The width, the peak radial position, and the intensity of the profiles are reasonably well reproduced by our simple model. However, the brightnesses of the profile central parts are overestimated in the three near-IR bands. As suggested by \citet{Foster2006}, this might be due to the simple density distribution chosen here, which may not be representative of the densest parts of the cloud. This might also be a size effect due to an increase in the number of big grains/monomers constituting the aggregates when the local density increases. Indeed, the \citet{Koehler2015} model considers only aggregates made of 4 monomers. \citet{Koehler2012} showed that further increasing the number of monomers in an aggregate grain does not change significantly the resulting emission in the far-IR/submm. However, increasing the number of monomers should modify the peak position of the scattering spectrum by shifting it towards slightly longer wavelengths, thus decreasing the scattering in the near-IR.

The two examples chosen here to illustrate the ability of THEMIS to reproduce the cloudshine observations are only well fitted by a CMM+AMMI grain mixture and not by a CMM+AMM mixture.  As written in Sect.~\ref{coreshine_cloudshine}, near-IR scattering is indeed more sensitive to the presence of an ice mantle than mid-IR scattering. Without this additional mantle, the scattering spectrum peaks at shorter wavelength (Fig.~\ref{Fig6}) leading to J, H, and K colour ratios disagreeing with the observations made by \citet{Malinen2013} and \citet{Foster2006}. The need for an ice mantle coating the aggregate grains complies with the findings of \citet{Andersen2014}, who showed a correlation between the dust scattering efficiency and the H$_2$O abundance in the Lupus IV molecular cloud complex.

\section{Discussion and conclusions}
\label{conclusions}

This study demonstrates the ability of our global evolutionary dust modelling approach THEMIS \citep{Jones2013, Koehler2014, Ysard2015, Koehler2015} to explain C-shine observations in the near- and mid-IR. The most important conclusion here is that with a combination of a-C:H mantle formation and low-level coagulation into aggregates we can self-consistently explain the observed C-shine and far-IR/submm emission towards dense starless clouds \citep{Koehler2015}. This combination also provides a tentative answer to two questions raised by previous studies about C-shine.

The first issue concerns the grain sizes required to explain the observations, a point recently addressed by \citet{Steinacker2015}. Most of the pioneering studies required that a large fraction of the dust mass was included in very big grains with sizes up to 1~$\mu$m \citep{Pagani2010, Steinacker2010, Andersen2013, Steinacker2014b}, whereas the calculation of the corresponding coagulation timescales according to the cloud densities and turbulence levels were discussed as inconsistent with the possibility to reach significant amounts of such big grains in the size distribution \citep{Steinacker2014b}. The aggregates as defined by \citet{Koehler2015} do not suffer from this inconsistency. Indeed, the accretion of H-rich carbon as a second mantle allows for a large increase in the scattering efficiency without the need for increasing the grain volume by more than a factor of $\sim 5$ (see paper I for a complete description of the grain optical properties and \citet{Koehler2012} for coagulation timescale calculations).

The second issue concerns the fact that inside a given region, coreshine is not detected in all the dense clouds observed by \citet{Paladini2014} and \citet{Lefevre2014} and that the proportion of clouds exhibiting coreshine varies from one region to another. For instance, 75\% of the dense clouds detected in Taurus exhibit coreshine, whereas in most other regions the proportion is closer to 50\% (such as Cepheus, Chamaeleon, and Musca)\footnote{We only discuss the detection of coreshine as, to our knowledge, there are no cloudshine large-scale study.}. On the contrary, there are for instance very few detections in the Orion region. In THEMIS, most of the scattering efficiency originates in the accretion of an a-C:H mantle. This leads to three possible explanations for the absence of detectable coreshine. The first explanation is related to the amount of carbon available in the gas phase. The abundance used by \citet{Koehler2015} relies on the highest C depletion measurements made by \citet{Parvathi2012} towards regions with $N_{\rm H} \geqslant 2 \times 10^{21}$~H/cm$^2$. \citet{Parvathi2012} highlighted the variability in the carbon depletion in dust depending on the line-of sight. Thus, there may be clouds were the amount of carbon available for a-C:H mantle formation is smaller or even close to zero: such regions would be populated with aggregates with a thinner H-rich carbon mantle or no second mantle at all and thus exhibit very little or no coreshine emission. A second explanation is related to the stability of H-rich carbon in the ISM, which depends strongly on the radiation field intensity to local density ratio \citep{Godard2011, Faraday2014}. In low density regions \citep[according to][$A_{\rm V} \leqslant 0.7$ for the standard ISRF]{Faraday2014}, UV photons are responsible for causing the photo-dissociation of CH bonds, a-C:H $\rightarrow$ a-C. In transition regions from diffuse ISM to dense clouds \citep[][$0.7 \leqslant A_{\rm V} \leqslant 1.2$ for the standard ISRF]{Faraday2014}, better shielded from UV photons and where the amount of hydrogen is significantly higher, H-poor carbon can be transformed into H-rich carbon through H atom incorporation, a-C $\rightarrow$ a-C:H. Similarly, carbon accreted from the gas phase in these transition regions is likely to be and stay H-rich. Then, in the dense molecular clouds, most of the hydrogen is in molecular form and thus not available to produce a-C:H mantles on the grains. However, this matches more or less the density at which ice mantles start to accrete on the grains, which would partly protect a-C:H layers formed earlier \citep[][and references therein]{Godard2011}. The stability and hydrogenation degree of a-C:H, as well as the exact values of $A_{\rm V}$ thresholds, are both dependent on the time scale and UV field intensity. The resulting a-C $\leftrightarrow$ a-C:H delicate balance could explain why in a quiet region such as Taurus most of the clouds exhibit coreshine, whereas in Orion, where on average the radiation field intensity and hardness are much higher, most clouds do not. A third explanation is related to the age and/or density of the clouds. In a young cloud, where dust growth is not advanced, or in an intermediate density cloud ($\rho_C \sim$~a few $10^3$~H/cm$^3$), the dust population may be dominated by CMM grains instead of AMM(I) dust. Such clouds would be as bright in the IRAC 8~$\mu$m band as in the two IRAC bands at 3.6 and 4.5~$\mu$m, thus not matching the selection criteria defined by \citet{Pagani2010} and \citet{Lefevre2014} and would be classified as ``no coreshine" clouds.

To really understand the variations in the observed C-shine, it will be required to fully quantify the composition and evolution of a-C(:H) mantles in the transition from the diffuse ISM to dense molecular clouds. This can only be done by combining carbon depletion measurements with detailed modelling of the dust emission from the near-IR to the submm. The equilibrium composition of carbon mantles will impact grain growth timescales, grain surface chemistry, as well as cloud mass estimates. In clouds harbouring young stellar objects (YSOs), the YSO influence on the dust exact composition will have to be reckoned before THEMIS can be applied in such regions.

\acknowledgements{We thank the anonymous referee for a very careful reading of this manuscript and useful remarks. This research was, in part, made possible through the EU FP7 funded project DustPedia (grant No. 606847). We would like to thank C. Lef\`evre for interesting discussions and for making all the coreshine data available. We are also grateful to M. Juvela for the 3D Monte-Carlo radiative transfer code CRT.}

\bibliography{biblio}

\begin{thebibliography}{63}
\expandafter\ifx\csname natexlab\endcsname\relax\def\natexlab#1{#1}\fi

\bibitem[{{Alata} {et~al.}(2014){Alata}, {Cruz-Diaz}, {Mu{\~n}oz Caro}, \&
  {Dartois}}]{Alata2014}
{Alata}, I., {Cruz-Diaz}, G.~A., {Mu{\~n}oz Caro}, G.~M., \& {Dartois}, E.
  2014, \aap, 569, A119

\bibitem[{{Alves de Oliveira} {et~al.}(2014){Alves de Oliveira}, {Schneider},
  {Mer{\'{\i}}n}, {Prusti}, {Ribas}, {Cox}, {Vavrek}, {K{\"o}nyves},
  {Arzoumanian}, {Puga}, {Pilbratt}, {K{\'o}sp{\'a}l}, {Andr{\'e}}, {Didelon},
  {Men'shchikov}, {Royer}, {Waelkens}, {Bontemps}, {Winston}, \&
  {Spezzi}}]{AlvesDeOliveira2014}
{Alves de Oliveira}, C., {Schneider}, N., {Mer{\'{\i}}n}, B., {et~al.} 2014,
  \aap, 568, A98

\bibitem[{{Andersen} {et~al.}(2013){Andersen}, {Steinacker}, {Thi}, {Pagani},
  {Bacmann}, \& {Paladini}}]{Andersen2013}
{Andersen}, M., {Steinacker}, J., {Thi}, W.-F., {et~al.} 2013, \aap, 559, A60

\bibitem[{{Andersen} {et~al.}(2014){Andersen}, {Thi}, {Steinacker}, \&
  {Tothill}}]{Andersen2014}
{Andersen}, M., {Thi}, W.-F., {Steinacker}, J., \& {Tothill}, N. 2014, \aap,
  568, L3

\bibitem[{{Arzoumanian} {et~al.}(2011){Arzoumanian}, {Andr{\'e}}, {Didelon},
  {K{\"o}nyves}, {Schneider}, {Men'shchikov}, {Sousbie}, {Zavagno}, {Bontemps},
  {di Francesco}, {Griffin}, {Hennemann}, {Hill}, {Kirk}, {Martin}, {Minier},
  {Molinari}, {Motte}, {Peretto}, {Pezzuto}, {Spinoglio}, {Ward-Thompson},
  {White}, \& {Wilson}}]{Arzoumanian2011}
{Arzoumanian}, D., {Andr{\'e}}, P., {Didelon}, P., {et~al.} 2011, \aap, 529, L6

\bibitem[{{Bernard} {et~al.}(1999){Bernard}, {Abergel}, {Ristorcelli}, {Pajot},
  {Torre}, {Boulanger}, {Giard}, {Lagache}, {Serra}, {Lamarre}, {Puget},
  {Lepeintre}, \& {Cambr{\'e}sy}}]{Bernard1999}
{Bernard}, J.~P., {Abergel}, A., {Ristorcelli}, I., {et~al.} 1999, \aap, 347,
  640

\bibitem[{{Boogert} {et~al.}(2015){Boogert}, {Gerakines}, \&
  {Whittet}}]{Boogert2015}
{Boogert}, A.~C.~A., {Gerakines}, P.~A., \& {Whittet}, D.~C.~B. 2015, \araa,
  53, 541

\bibitem[{{Compi{\`e}gne} {et~al.}(2011){Compi{\`e}gne}, {Verstraete}, {Jones},
  {Bernard}, {Boulanger}, {Flagey}, {Le Bourlot}, {Paradis}, \&
  {Ysard}}]{Compiegne2011}
{Compi{\`e}gne}, M., {Verstraete}, L., {Jones}, A., {et~al.} 2011, \aap, 525,
  A103

\bibitem[{{Dartois}(2006)}]{Dartois2006}
{Dartois}, E. 2006, \aap, 445, 959

\bibitem[{{Fanciullo} {et~al.}(2015){Fanciullo}, {Guillet}, {Aniano}, {Jones},
  {Ysard}, {Miville-Desch{\^e}nes}, {Boulanger}, \&
  {K{\"o}hler}}]{Fanciullo2015}
{Fanciullo}, L., {Guillet}, V., {Aniano}, G., {et~al.} 2015, \aap, 580, A136

\bibitem[{{Flagey} {et~al.}(2009){Flagey}, {Noriega-Crespo}, {Boulanger},
  {Carey}, {Brooke}, {Falgarone}, {Huard}, {McCabe}, {Miville-Desch{\^e}nes},
  {Padgett}, {Paladini}, \& {Rebull}}]{Flagey2009}
{Flagey}, N., {Noriega-Crespo}, A., {Boulanger}, F., {et~al.} 2009, \apj, 701,
  1450

\bibitem[{{Foster} \& {Goodman}(2006)}]{Foster2006}
{Foster}, J.~B. \& {Goodman}, A.~A. 2006, \apjl, 636, L105

\bibitem[{{Godard} {et~al.}(2011){Godard}, {F{\'e}raud}, {Chabot},
  {Carpentier}, {Pino}, {Brunetto}, {Duprat}, {Engrand}, {Br{\'e}chignac},
  {D'Hendecourt}, \& {Dartois}}]{Godard2011}
{Godard}, M., {F{\'e}raud}, G., {Chabot}, M., {et~al.} 2011, \aap, 529, A146

\bibitem[{{Henyey} \& {Greenstein}(1941)}]{HG1941}
{Henyey}, L.~G. \& {Greenstein}, J.~L. 1941, \apj, 93, 70

\bibitem[{{Jones}(2012{\natexlab{a}})}]{Jones2012a}
{Jones}, A.~P. 2012{\natexlab{a}}, \aap, 540, A1

\bibitem[{{Jones}(2012{\natexlab{b}})}]{Jones2012b}
{Jones}, A.~P. 2012{\natexlab{b}}, \aap, 540, A2

\bibitem[{{Jones}(2012{\natexlab{c}})}]{Jones2012c}
{Jones}, A.~P. 2012{\natexlab{c}}, \aap, 542, A98

\bibitem[{{Jones} {et~al.}(2013){Jones}, {Fanciullo}, {K\"ohler}, {Verstraete},
  {Guillet}, {Bocchio}, \& {Ysard}}]{Jones2013}
{Jones}, A.~P., {Fanciullo}, L., {K\"ohler}, M., {et~al.} 2013, \aap, 558, A62

\bibitem[{{Jones} {et~al.}(2015){Jones}, {Koehler}, {Ysard}, {Dartois},
  {Godard}, \& {Gavilan}}]{Jones2015}
{Jones}, A.~P., {Koehler}, M., {Ysard}, N., {et~al.} 2015, A\&A, submitted

\bibitem[{{Jones} {et~al.}(2014){Jones}, {Ysard}, {K\"ohler}, {Fanciullo},
  {Bocchio}, {Micelotta}, {Verstraete}, \& {Guillet}}]{Faraday2014}
{Jones}, A.~P., {Ysard}, N., {K\"ohler}, M., {et~al.} 2014, Faraday Discuss.,
  168

\bibitem[{{Juvela}(2005)}]{Juvela2005}
{Juvela}, M. 2005, \aap, 440, 531

\bibitem[{{Juvela} {et~al.}(2008){Juvela}, {Pelkonen}, {Padoan}, \&
  {Mattila}}]{Juvela2008}
{Juvela}, M., {Pelkonen}, V.-M., {Padoan}, P., \& {Mattila}, K. 2008, \aap,
  480, 445

\bibitem[{{Juvela} {et~al.}(2009){Juvela}, {Pelkonen}, \&
  {Porceddu}}]{Juvela2009}
{Juvela}, M., {Pelkonen}, V.-M., \& {Porceddu}, S. 2009, \aap, 505, 663

\bibitem[{{Juvela} {et~al.}(2012){Juvela}, {Ristorcelli}, {Pagani}, {Doi},
  {Pelkonen}, {Marshall}, {Bernard}, {Falgarone}, {Malinen}, {Marton},
  {McGehee}, {Montier}, {Motte}, {Paladini}, {T{\'o}th}, {Ysard}, {Zahorecz},
  \& {Zavagno}}]{Juvela2012}
{Juvela}, M., {Ristorcelli}, I., {Pagani}, L., {et~al.} 2012, \aap, 541, A12

\bibitem[{{K\"ohler} {et~al.}(2014){K\"ohler}, {Jones}, \&
  {Ysard}}]{Koehler2014}
{K\"ohler}, M., {Jones}, A., \& {Ysard}, N. 2014, \aap, 565, L9

\bibitem[{{K\"ohler} {et~al.}(2012){K\"ohler}, {Stepnik}, {Jones}, {Guillet},
  {Abergel}, {Ristorcelli}, \& {Bernard}}]{Koehler2012}
{K\"ohler}, M., {Stepnik}, B., {Jones}, A.~P., {et~al.} 2012, \aap, 548, A61

\bibitem[{{K\"ohler} {et~al.}(2015){K\"ohler}, {Ysard}, \&
  {Jones}}]{Koehler2015}
{K\"ohler}, M., {Ysard}, N., \& {Jones}. 2015, \aap

\bibitem[{{Kov{\'a}cs} {et~al.}(2004){Kov{\'a}cs}, {Mall}, {Bizenberger},
  {Baumeister}, \& {R{\"o}ser}}]{Kovacs2004}
{Kov{\'a}cs}, Z., {Mall}, U., {Bizenberger}, P., {Baumeister}, H., \&
  {R{\"o}ser}, H.-J. 2004, in Society of Photo-Optical Instrumentation
  Engineers (SPIE) Conference Series, Vol. 5499, Optical and Infrared Detectors
  for Astronomy, ed. J.~D. {Garnett} \& J.~W. {Beletic}, 432--441

\bibitem[{{Lagache} {et~al.}(1998){Lagache}, {Abergel}, {Boulanger}, \&
  {Puget}}]{Lagache1998}
{Lagache}, G., {Abergel}, A., {Boulanger}, F., \& {Puget}, J.-L. 1998, \aap,
  333, 709

\bibitem[{{Laureijs} {et~al.}(1991){Laureijs}, {Clark}, \&
  {Prusti}}]{Laureijs1991}
{Laureijs}, R.~J., {Clark}, F.~O., \& {Prusti}, T. 1991, \apj, 372, 185

\bibitem[{{Lef{\`e}vre} {et~al.}(2014){Lef{\`e}vre}, {Pagani}, {Juvela},
  {Paladini}, {Lallement}, {Marshall}, {Andersen}, {Bacmann}, {McGehee},
  {Montier}, {Noriega-Crespo}, {Pelkonen}, {Ristorcelli}, \&
  {Steinacker}}]{Lefevre2014}
{Lef{\`e}vre}, C., {Pagani}, L., {Juvela}, M., {et~al.} 2014, \aap, 572, A20

\bibitem[{{Lehtinen} \& {Mattila}(1996)}]{Lehtinen1996}
{Lehtinen}, K. \& {Mattila}, K. 1996, \aap, 309, 570

\bibitem[{{Malinen} {et~al.}(2013){Malinen}, {Juvela}, {Pelkonen}, \&
  {Rawlings}}]{Malinen2013}
{Malinen}, J., {Juvela}, M., {Pelkonen}, V.-M., \& {Rawlings}, M.~G. 2013,
  \aap, 558, A44

\bibitem[{{Malinen} {et~al.}(2012){Malinen}, {Juvela}, {Rawlings},
  {Ward-Thompson}, {Palmeirim}, \& {Andr{\'e}}}]{Malinen2012}
{Malinen}, J., {Juvela}, M., {Rawlings}, M.~G., {et~al.} 2012, \aap, 544, A50

\bibitem[{{Mathis} {et~al.}(1983){Mathis}, {Mezger}, \& {Panagia}}]{Mathis1983}
{Mathis}, J.~S., {Mezger}, P.~G., \& {Panagia}, N. 1983, \aap, 128, 212

\bibitem[{{Mattila}(1970{\natexlab{a}})}]{Mattila1970b}
{Mattila}, K. 1970{\natexlab{a}}, \aap, 9, 53

\bibitem[{{Mattila}(1970{\natexlab{b}})}]{Mattila1970a}
{Mattila}, K. 1970{\natexlab{b}}, \aap, 8, 273

\bibitem[{{Meng} {et~al.}(2013){Meng}, {Wu}, \& {Liu}}]{Meng2013}
{Meng}, F., {Wu}, Y., \& {Liu}, T. 2013, \apjs, 209, 37

\bibitem[{{Nakajima} {et~al.}(2003){Nakajima}, {Nagata}, {Sato}, {Nagayama},
  {Nagashima}, {Kato}, {Kurita}, {Kawai}, {Tamura}, {Nakaya}, \&
  {Sugitani}}]{Nakajima2003}
{Nakajima}, Y., {Nagata}, T., {Sato}, S., {et~al.} 2003, \aj, 125, 1407

\bibitem[{{Narayanan} {et~al.}(2008){Narayanan}, {Heyer}, {Brunt}, {Goldsmith},
  {Snell}, \& {Li}}]{Narayanan2008}
{Narayanan}, G., {Heyer}, M.~H., {Brunt}, C., {et~al.} 2008, \apjs, 177, 341

\bibitem[{{Ormel} {et~al.}(2009){Ormel}, {Paszun}, {Dominik}, \&
  {Tielens}}]{Ormel2009}
{Ormel}, C.~W., {Paszun}, D., {Dominik}, C., \& {Tielens}, A.~G.~G.~M. 2009,
  \aap, 502, 845

\bibitem[{{Ossenkopf}(1993)}]{Ossenkopf1993}
{Ossenkopf}, V. 1993, \aap, 280, 617

\bibitem[{{Padoan} {et~al.}(2002){Padoan}, {Cambr{\'e}sy}, \&
  {Langer}}]{Padoan2002}
{Padoan}, P., {Cambr{\'e}sy}, L., \& {Langer}, W. 2002, \apjl, 580, L57

\bibitem[{{Pagani} {et~al.}(2010){Pagani}, {Steinacker}, {Bacmann}, {Stutz}, \&
  {Henning}}]{Pagani2010}
{Pagani}, L., {Steinacker}, J., {Bacmann}, A., {Stutz}, A., \& {Henning}, T.
  2010, Science, 329, 1622

\bibitem[{{Paladini}(2014)}]{Paladini2014}
{Paladini}, R. 2014, Astrophysics and Space Science Proceedings, 36, 299

\bibitem[{{Parvathi} {et~al.}(2012){Parvathi}, {Sofia}, {Murthy}, \&
  {Babu}}]{Parvathi2012}
{Parvathi}, V.~S., {Sofia}, U.~J., {Murthy}, J., \& {Babu}, B.~R.~S. 2012,
  \apj, 760, 36

\bibitem[{{Qian} {et~al.}(2012){Qian}, {Li}, \& {Goldsmith}}]{Qian2012}
{Qian}, L., {Li}, D., \& {Goldsmith}, P.~F. 2012, \apj, 760, 147

\bibitem[{{Ridderstad} {et~al.}(2006){Ridderstad}, {Juvela}, {Lehtinen},
  {Lemke}, \& {Liljestr{\"o}m}}]{Ridderstad2006}
{Ridderstad}, M., {Juvela}, M., {Lehtinen}, K., {Lemke}, D., \&
  {Liljestr{\"o}m}, T. 2006, \aap, 451, 961

\bibitem[{{Roy} {et~al.}(2013){Roy}, {Martin}, {Polychroni}, {Bontemps},
  {Abergel}, {Andr{\'e}}, {Arzoumanian}, {Di Francesco}, {Hill}, {Konyves},
  {Nguyen-Luong}, {Pezzuto}, {Schneider}, {Testi}, \& {White}}]{Roy2013}
{Roy}, A., {Martin}, P.~G., {Polychroni}, D., {et~al.} 2013, \apj, 763, 55

\bibitem[{{Schnee} {et~al.}(2010){Schnee}, {Enoch}, {Noriega-Crespo}, {Sayers},
  {Terebey}, {Caselli}, {Foster}, {Goodman}, {Kauffmann}, {Padgett}, {Rebull},
  {Sargent}, \& {Shetty}}]{Schnee2010}
{Schnee}, S., {Enoch}, M., {Noriega-Crespo}, A., {et~al.} 2010, \apj, 708, 127

\bibitem[{{Smith} {et~al.}(1989){Smith}, {Sellgren}, \& {Tokunaga}}]{Smith1989}
{Smith}, R.~G., {Sellgren}, K., \& {Tokunaga}, A.~T. 1989, \apj, 344, 413

\bibitem[{{Steinacker} {et~al.}(2014{\natexlab{a}}){Steinacker}, {Andersen},
  {Thi}, \& {Bacmann}}]{Steinacker2014a}
{Steinacker}, J., {Andersen}, M., {Thi}, W.-F., \& {Bacmann}, A.
  2014{\natexlab{a}}, \aap, 563, A106

\bibitem[{{Steinacker} {et~al.}(2015){Steinacker}, {Andersen}, {Thi},
  {Paladini}, {Juvela}, {Bacmann}, {Pelkonen}, {Pagani}, {Lef{\`e}vre},
  {Henning}, \& {Noriega-Crespo}}]{Steinacker2015}
{Steinacker}, J., {Andersen}, M., {Thi}, W.-F., {et~al.} 2015, ArXiv e-prints

\bibitem[{{Steinacker} {et~al.}(2014{\natexlab{b}}){Steinacker}, {Ormel},
  {Andersen}, \& {Bacmann}}]{Steinacker2014b}
{Steinacker}, J., {Ormel}, C.~W., {Andersen}, M., \& {Bacmann}, A.
  2014{\natexlab{b}}, \aap, 564, A96

\bibitem[{{Steinacker} {et~al.}(2010){Steinacker}, {Pagani}, {Bacmann}, \&
  {Guieu}}]{Steinacker2010}
{Steinacker}, J., {Pagani}, L., {Bacmann}, A., \& {Guieu}, S. 2010, \aap, 511,
  A9

\bibitem[{{Stepnik} {et~al.}(2003){Stepnik}, {Abergel}, {Bernard}, {Boulanger},
  {Cambr{\'e}sy}, {Giard}, {Jones}, {Lagache}, {Lamarre}, {Meny}, {Pajot}, {Le
  Peintre}, {Ristorcelli}, {Serra}, \& {Torre}}]{Stepnik2003}
{Stepnik}, B., {Abergel}, A., {Bernard}, J.-P., {et~al.} 2003, \aap, 398, 551

\bibitem[{{Struve}(1937)}]{Struve1937}
{Struve}, O. 1937, \apj, 85, 194

\bibitem[{{Struve} \& {Elvey}(1936)}]{Struve1936}
{Struve}, O. \& {Elvey}, C.~T. 1936, \apj, 83, 162

\bibitem[{{Witt}(1968)}]{Witt1968}
{Witt}, A.~N. 1968, \apj, 152, 59

\bibitem[{{Witt} {et~al.}(1994){Witt}, {Lindell}, {Block}, \&
  {Evans}}]{Witt1994}
{Witt}, A.~N., {Lindell}, R.~S., {Block}, D.~L., \& {Evans}, R. 1994, \apj,
  427, 227

\bibitem[{{Ysard} {et~al.}(2013){Ysard}, {Abergel}, {Ristorcelli}, {Juvela},
  {Pagani}, {K{\"o}nyves}, {Spencer}, {White}, \& {Zavagno}}]{Ysard2013}
{Ysard}, N., {Abergel}, A., {Ristorcelli}, I., {et~al.} 2013, \aap, 559, A133

\bibitem[{{Ysard} {et~al.}(2012){Ysard}, {Juvela}, {Demyk}, {Guillet},
  {Abergel}, {Bernard}, {Malinen}, {M{\'e}ny}, {Montier}, {Paradis},
  {Ristorcelli}, \& {Verstraete}}]{Ysard2012}
{Ysard}, N., {Juvela}, M., {Demyk}, K., {et~al.} 2012, \aap, 542, A21

\bibitem[{{Ysard} {et~al.}(2015){Ysard}, {K\"ohler}, {Jones},
  {Miville-Desch{\^e}nes}, {Abergel}, \& {Fanciullo}}]{Ysard2015}
{Ysard}, N., {K\"ohler}, M., {Jones}, A., {et~al.} 2015, \aap, 577, A110

\end{thebibliography}

\end{document}